\RequirePackage{fix-cm}

\documentclass[twocolumn]{svjour3}        

\smartqed  
\usepackage[final]{graphicx}

\usepackage{multirow}

\usepackage{natbib}
\usepackage{doi}
\usepackage{url}
\usepackage{breakurl}
\usepackage{textcomp}
\usepackage[labelsep=space]{caption}
\usepackage{subcaption}
\usepackage{float}
\usepackage{amssymb}
\usepackage{amsmath}
\usepackage{mathtools}
\usepackage{setspace}
\usepackage[dvipsnames]{xcolor}
\usepackage{array}
\usepackage{tabularx}
\hypersetup{hidelinks}

\usepackage[T1]{fontenc}

\usepackage[inkscapelatex=false]{svg}

\usepackage[latin1]{inputenc}
\usepackage{tikz}
\usetikzlibrary{shapes.geometric,arrows,fit,positioning}

\renewcommand*{\doi}[1]{\href{http://dx.doi.org/#1}{http://dx.doi.org/#1}}

\begin{document} \sloppy

\section*{Graphical Abstract}
\begin{figure}[h!]
\centering
    \includegraphics{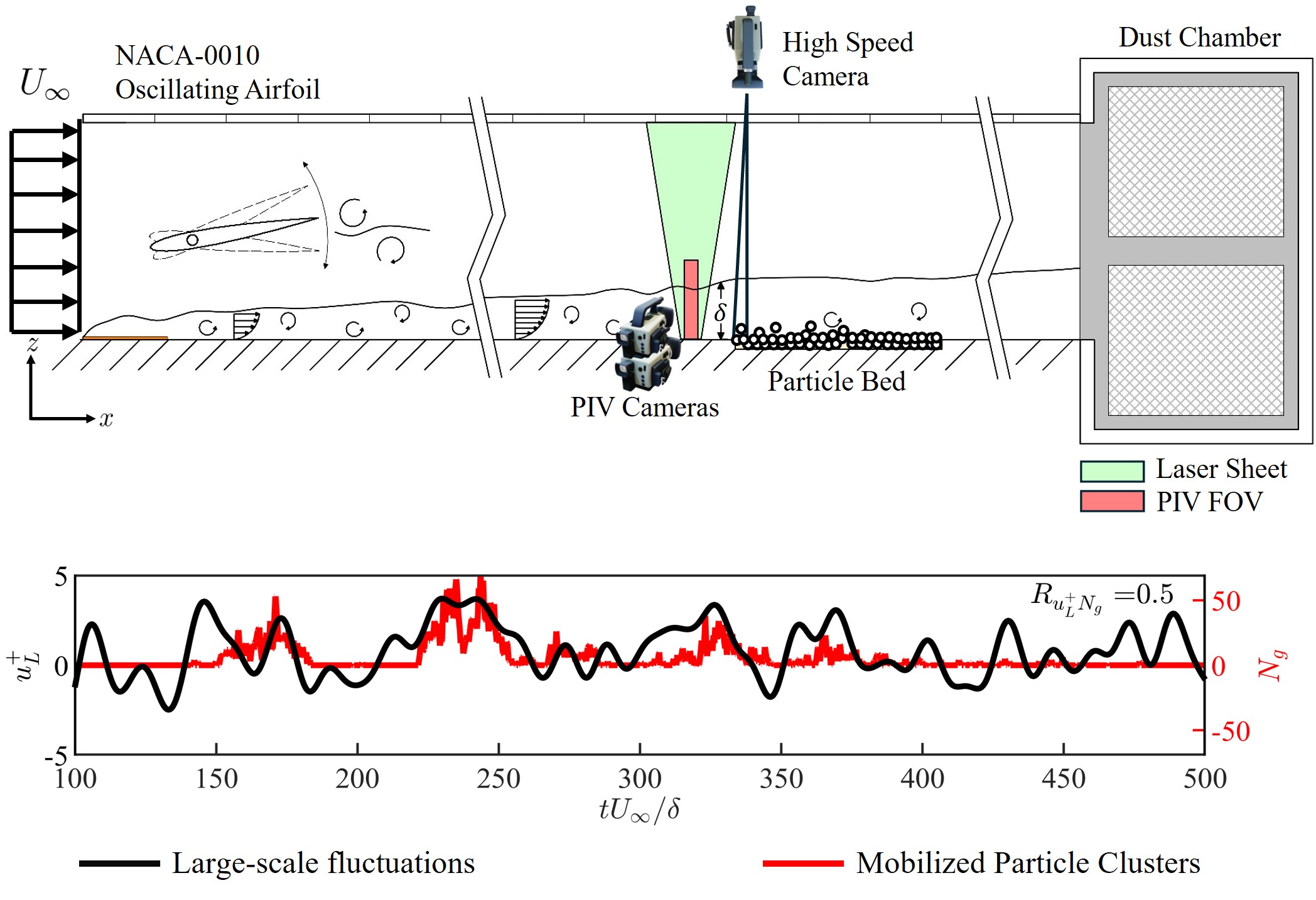}
\end{figure}

\title{An Experimental Framework to Study Turbulence Induced Particle Mobilization}

\author{Vaishak Thiruvenkitam \and Robert H. Bryan II \and Zheng Zhang \and Ebenezer P. Gnanamanickam}

\institute{Ebenezer P. Gnanamanickam \at
              Associate Professor,\\
              Department of Aerospace Engineering, \\
              Embry-Riddle Aeronautical University,\\
              Daytona Beach, FL, USA.\\
              Tel.: +1 386-226-7532\\
              \email{gnanamae@erau.edu}  
}

\date{Received: date / Accepted: date}

\maketitle

\begin{abstract}

An experimental framework was developed to study the initiation of particle mobilization in a laboratory setting. Large and heavy particles mobilized by a turbulent, gaseous carrier-phase were considered. An airfoil oscillated in the free-stream, generating a tonal free-stream disturbance that perturbed a turbulent boundary layer. The flow developing behind this forced flow was characterized using hot-wire anemometry and particle image velocimetry (PIV). Downstream of the oscillating airfoil mechanism, the turbulent boundary layer responded to the forcing in the form of excess energy at the forcing scale. The signature of this forcing scale was observed to span the entire wall-normal extent of the flow, extending all the way down to the wall. The size of this flow scale was shown to be controlled by changing the frequency of oscillation, while the energy in this flow scale was controlled via the amplitude of oscillation. Demonstrative measurements are presented in which this forced carrier-phase flow mobilized a particle-phase on a particle bed. A PIV-based approach was used to measure the initiation of particle motion as well as the incoming carrier-phase velocity field. The particles on the bed were mobilized ``on-demand'' by the deflection of the airfoil. Consistent with prior work, it was observed that particle mobilization was correlated with the large-scale flow structures of the carrier-phase.

\end{abstract}

\section{Introduction}

The entrainment of solid particles by air is a common occurrence, with the most familiar instance being the transport of aeolian particles (such as sand, dust, and ice particles) by the Earth's boundary layer. Such transport of particles within a planetary boundary layer is referred to broadly as aeolian transport. A particle on a particle bed, under the influence of the Earth's atmospheric boundary layer, would be dislodged from its static position if the total forces (and/or moments) on the particle reached some threshold that overcame the forces (and/or moments) holding the particle at rest~\citep{Bagnold1971a,Kok2012a,Phatz2020a}. While these aeolian processes occur over a range of particle diameters, the present work focuses on large particles, i.e., particle-phase with particle diameter $d_p>100$~\textmu m. In addition, these particles are mobilized by air within a turbulent boundary layer, referred to hereon as the carrier-phase. In this regime, apart from the aerodynamic forces acting on these particles, gravity as well as contact forces with neighboring particles are significant~\citep{Duran2011a}. Further, when particles are mobilized from a bed, they either roll on or hop and skip on the surface, with the latter motion referred to as saltation. For the rest of this work, the process of particles being initiated into either of these modes of motion is collectively referred to as particle mobilization. 

The flow conditions leading to the initiation of particle motion on the surface of particle beds are critical for natural planetary processes, particularly the formation of surface patterns such as dunes~\citep{Bagnold1971a}. In an engineering context, the motion of vehicles, such as mining vehicles and aircraft, generates energetic, large-scale carrier-phase eddies that caused the mobilization of large particles. Operating such vehicles in environments where particles are mobilized could lead to the degradation of sensory inputs (both human and instrument) because of the presence of the mobilized particle-phase~\citep{Rauleder2014a}. Another related but novel flow scenario involves the use of rotorcraft (and the large-scale structures of its wake) to intentionally mobilize particles in an extraterrestrial environment in a controlled manner~\citep{Rabinovitch2021a}. Such an approach provides a means to understand extraterrestrial aeolian processes that might otherwise be intractable. Thus, understanding particle mobilization (and its relationship with the carrier-phase flow scales) has relevance not only to the fundamental processes underpinning these complex natural flow fields but also for the design of engineering systems intended to operate in these environments. 

There are several excellent reviews that focus on the various approaches used to predict the initiation of particle mobilization, and readers are referred to these for a complete treatment~\citep{Duran2011a,Kok2012a,Yang2019a,Phatz2020a}. Here, drawing from these reviews, a brief summary of the current approaches is provided. By considering the balance of forces on a particle resting on a bed surface, \citet{Bagnold1971a} obtained an expression for the threshold friction velocity $u_\tau^t=\Theta\sqrt{\left(\rho_p-\rho_c\right)g d_p/\rho_c}$ above which a particle will undergo mobilization. Here, $\rho_p$ and $\rho_c$ are the densities of the particle and carrier-phase, respectively, $g$ denotes the acceleration due to gravity, and $\Theta$ is a constant. This formulation stands as the earliest and most prominently used measure to predict the initiation of particle mobilization and is referred to as the static threshold. This condition is frequently expressed via the non-dimensional number $S=\sqrt{\Theta}=\tau^t/\left[\left(\rho_p-\rho_c\right)gd_p\right]$, referred to as the Shields number or parameter~\citep{Phatz2020a,Valance2015a,Kok2012a,Duran2011a}. Here, $\tau^t$ is the threshold shear stress at the surface of the particle bed. 

This shear stress at the bed surface increases as the flow speed increases and once some experimentally determined critical threshold $\tau^t$ is reached, particles are considered to be mobilized. \citet{Bagnold1971a} determined that for air, in the context of the static threshold, $\Theta\approx0.1$. Note, in the case of water $\Theta\approx 0.2$, a difference that points to differences in particle mobilization with varying density ratios $\rho_p/\rho_c$~\citep{Burr2020a}. Various other, more sophisticated estimates were subsequently used to predict the threshold friction velocity; see, for example, ~\citet{Iversen1982a,Shao2000a,Kok2006a}. In general, experimentally verified measures showed that the threshold friction velocity $u_\tau^t$ decreased as the particle diameter increased, reached a minimum at around $d_p=75$ \textmu m, and increased as the particle diameter became larger. Closely related to the concept of a static threshold is the dynamic threshold. Once particles are moved from their position of rest on a particle bed, they roll or saltate and mobilize other particles at a shear velocity, referred to as the dynamic threshold $u_\tau^d$, that is lower than the static threshold $u_\tau^t$. This then implied a lower $\Theta$ and indeed \citet{Bagnold1971a} estimated that in the case of the dynamic threshold $\Theta\approx0.08$.

Any shear stress-based measure that predicted particle mobilization relied on a time-averaged quantity. However, it is clear to even a casual observer that the mobilization of particles, for instance, dust or snow particles by wind, is spasmodic and correlated with wind gusts or flow structures of the carrier-phase. Indeed, \citet{Stout1997a} considered the intermittency of particle mobilization and showed that the period during which particles were mobilized represented a small fraction of the total time period under consideration. Hence, for a complete understanding of particle mobilization, it became necessary to characterize this process in the context of individual flow scales as opposed to a time-averaged measure. With the increased use of scale-resolving simulations, it is also necessary to predict particle mobilization locally and instantaneously for the development of particle scale mobilization models~\citep{Lee2012a}. 

Thus, over the last two decades, considerable focus has been placed on characterizing particle mobilization in terms of the fluctuating characteristics (or individual flow eddies) of the carrier-phase flow. \citet{Diplas2008a} and \citet{Valyrakis2010a} in a novel experiment used an electromagnet to impart an impulsive force to mobilize particles. Based on these experiments, they hypothesized that particles could be mobilized from a particle bed when subjected to an impulsive force that exceeded a threshold in magnitude. The force in this case would be the aerodynamic forces on the particle imparted by the turbulent carrier-phase eddies. In addition, these impulsive forces needed to persist over a certain time period to move particles from their pockets of rest. In this context, the combined works of \citet{Xuan1994a} and \citet{Xuan2004a} showed that as the turbulence intensity of the carrier-phase flow increased, there was a corresponding decrease in the static threshold. While the turbulence intensity itself is an integrated (across all flow scales) quantity, its influence points to the potential importance of the average energy of carrier-phase flow structures and the impulsive forces generated by them. In a similar vein, a work-based framework was also considered, wherein the turbulent structures of the carrier-phase worked on particles to dislodge them from rest~\citep{Lee2012a}. \citet{Valyrakis2013a} used an energy-based approach to consider particle mobilization, where particles were considered to be mobilized when two conditions were satisfied. The first condition is satisfied when the energy rate imparted by the carrier-phase eddies exceeded a threshold, while the second condition simultaneously required that the scale-size of the incoming flow eddy should be sufficiently large, above some threshold size. This energy-based approach was evaluated in the context of water erosion, via the use of an experimentally determined energy transfer coefficient, with satisfactory outcomes~\citep{Valyrakis2013a}. 

Two recent works conducted in flow fields where the carrier-phase was air have relevance to the present work. The first was an experiment performed in a wind tunnel where a cloth fluttered freely upstream of a sand bed~\citep{Zhang2022a}. The fluttering perturbed the flow, generating large-scale energetic flow structures that resulted in a variable shear at the bed surface. More importantly, these structures, while not changing the mean shear stress at the bed surface, mobilized significantly more particles when compared with the unperturbed flow. Since the unperturbed flow had very little energy at the scales forced by the fluttering cloth, it was concluded that increased particle mobilization was attributed to the increased energy in the flow scales associated with the fluttering. More recently, \citet{Valyrakis2025a} conducted wind tunnel experiments studying the mobilization of coarse ($d_p=40000$ \textmu m) very light ($\rho_p/\rho_c\approx67$) particles by the energetic structures of the carrier-phase. Through the use of the aforementioned energy transfer coefficient \citep{Valyrakis2013a} they demonstrated that the incipient motion of these particles was coupled with specific energetic carrier-phase flow structures. Several studies that considered water erosion (lower density ratio than air flows) also reported the correlation between the instantaneous energetic flow structures of the carrier-phase and particle mobilization~\citep{Wu2012b,Cameron2020a}. 

Hence, while significant progress has been made in understanding the interactions between the carrier-phase eddies and the particle bed, a complete understanding remains elusive, with several fundamental questions remaining unanswered. For example, for a given density ratio $\rho_p/\rho_c$ and particle Reynolds number $Re_p=u_\tau d_p/\nu$, does a specific carrier-phase eddy scale and energy exist that maximizes particle mobilization? Conversely, what range of flow scales likely minimizes the mobilization of particles of a given characteristic? Gaining such understanding requires a controlled study where particle-phase characteristics as well as carrier-phase eddy characteristics are systematically varied. Such studies are, however, challenging if not impossible to conduct in the field. Wind tunnel studies, on the other hand, present their own challenges, particularly concerning their relevance to field environments. Most notably are the challenges that arise due to the differences in Reynolds number between field campaigns and wind tunnel studies. As the Reynolds number of a turbulent boundary layer increases, the largest scales in the flow become larger, and the energy in these large-scales starts to increase~\citep{Hutchins2007b,Smits2011a}. Here, large-scales refer to scales of the order of the boundary layer thickness $\delta$ and larger, which in the atmospheric boundary layer are of the order of kilometers~\citep{Hutchins2012a}. At high Reynolds numbers, the large-scales are the significant energy-bearing eddies, in contrast with lower Reynolds number laboratory flows~\citep{Smits2011a}. Thus, laboratory flows may not have sufficient energy in the eddies of scale sizes that interact with large and heavy particles such as those considered presently.

To this end, the present work introduces a framework where energetic large-scales are introduced via an oscillating airfoil mechanism. While acting as a method to introduce controlled large-scales into the flow, the oscillating airfoil also introduced free-stream turbulence. Several excellent works characterized the turbulent boundary layer in the presence of broadband free-stream turbulence~\citep{Dogan2016a,Sharp2009a,Hancock1989a}. Depending on the level of free-stream turbulence, it was shown that the effects could penetrate deep into the boundary layer and modify the near-wall flow. In this work, essentially a tonal disturbance is introduced into the flow via the oscillating airfoil mechanism. Both the size of the introduced scale and the energy in the flow scales could be altered via the oscillation mechanism. A particle bed was subjected to these large-scale structures, and otherwise immobile particles were mobilized by the introduced scales. Using these demonstrative experiments, the entire experimental framework is characterized and presented herein. First, the entire experimental approach is described. Then the changes to the carrier-phase caused by the oscillating airfoil (hereon referred to as the forced flow) are characterized. Following this, the two-phase flow is considered, where particles on a particle bed are subjected to the forced flow, and demonstrative measurements are presented. Finally, the primary conclusions are summarized.

\section{Experimental Approach}
\label{sec:Approach}

\begin{figure*}[tpb]
    \centering
    \includegraphics[width=0.9\textwidth]{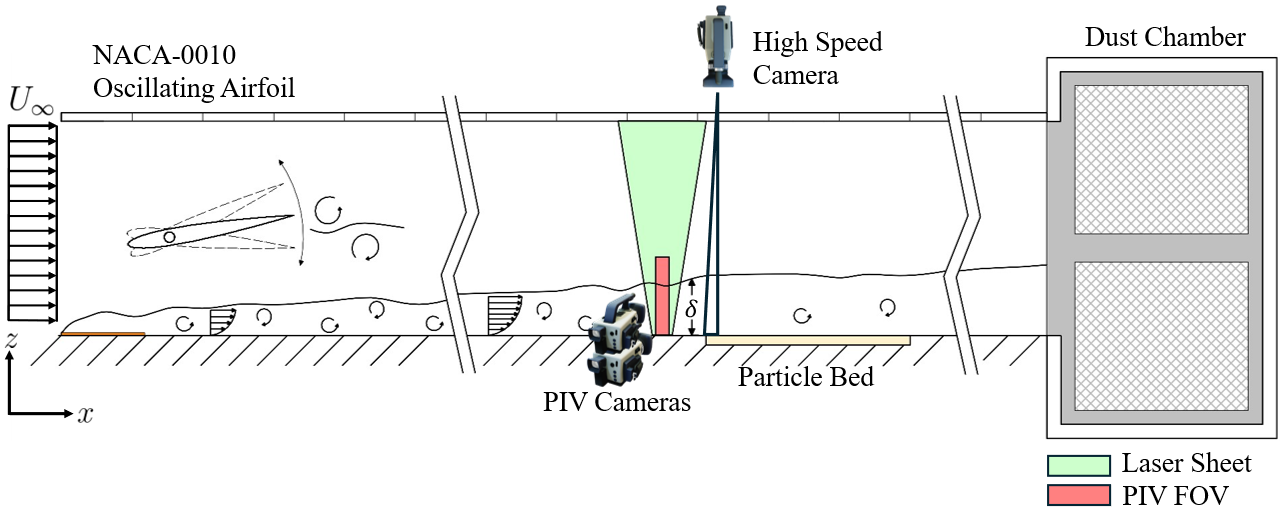}
    \caption{Schematic highlighting the salient features of the experimental framework. The red-shaded region represents the combined fields of view of the two PIV cameras. The green shaded region depicts the laser sheet. At the exit to the tunnel, a dust chamber collects the particles mobilized by the flow. Note that this schematic is not to scale}
    \label{fig:sch}
\end{figure*}

In order to systematically study the initiation of particle mobilization, it was imperative to develop a repeatable process to consistently introduce large-scale structures into the flow. Additionally, controlling the energy of these large-scale structures was necessary. This was accomplished using an oscillating airfoil mechanism, as shown in the schematics of Fig.~\ref{fig:sch} and Fig.~\ref{fig:OscillatingMechanism}. This section describes this experimental approach, including information on the facility and measurement techniques used. 

\subsection{Notation}

In the interest of clarity, the major notation used is summarized first. The streamwise and wall-normal ordinates are denoted by $x$ and $z$, respectively. The friction velocity is represented by $U_\tau$, the kinematic viscosity by $\nu$, the carrier-phase density by $\rho_c$, the free-stream velocity by $U_\infty$, and the boundary layer thickness by $\delta$. The average components of velocity in the streamwise and wall-normal directions are represented by $\overline{U}$ and $\overline{W}$, respectively, whereas their respective fluctuation components are denoted by $u$ and $w$. The streamwise flow scales are of wavelength $\lambda_x$ and wavenumber $k_x$. The oscillating airfoil introduced large-scale structures with a wavelength $\lambda_f$ while oscillating at a frequency $f_f$, with an amplitude of oscillation denoted by $A_f$. The time-varying phase of the airfoil is denoted by $\theta_{f}$, which is used as a reference for phase-locked calculations. The phase-locked streamwise velocity fluctuations are then denoted by $\tilde{u}$. Similarly, the ensemble average of conditional streamwise and wall-normal velocity fluctuations are $\hat{u}$ and $\hat{w}$ respectively. The superscript `+' denotes non-dimensionalization using viscous scales, where the characteristic length, velocity, and time scales are $U_{\tau}/\nu$, $U_{\tau}$, and $U_{\tau}^2/\nu$, respectively. The convection velocity is denoted by $U_{c}$. The particle-phase had particles of diameter $d_p$ and density $\rho_p$. Finally, $N_{g}$ denotes the mobile particle clusters mobilized at any given instant of time.

\subsection{Experimental Facility}
All experiments were conducted in the Embry-Riddle Aeronautical University's low-speed boundary layer wind tunnel (BLWT). The entrance to the test section of the facility measured 0.645~m in width and 0.355~m in height. A 60-grit sandpaper strip was placed at the entrance to the test section to transition the boundary layer, allowing a turbulent boundary layer to develop over a streamwise distance exceeding 4.8~m. The roof of the tunnel was adjusted to maintain a nominally zero-pressure-gradient boundary layer for all flow scenarios presented in this work. To contain the mobilized particles, an enclosure was constructed at the exit to the test section. Large portions of the enclosure were covered with stainless steel mesh to retain particles within the enclosure.

 All measurements were conducted at approximately 2.7~m downstream of the boundary-layer-trip. The tunnel can operate at free-stream velocities of over 15~ms$^{-1}$. However, all experiments presented in this work were carried out at a nominal free-stream velocity of $U_{\infty}\approx$~9~ms$^{-1}$. This free-stream velocity was chosen for the following reason. At the streamwise measurement location under consideration ($\approx$~2.7~m downstream of the trip), when a particle bed was introduced into the flow, the mean shear at the wall resulted in little to no natural particle mobilization, i.e., the particle bed was not depleted. Even a small increase in the free-stream velocity beyond this threshold velocity resulted in natural particle mobilization. This threshold condition was determined using repeated careful trials. This approach then allowed for the introduction of a controlled large-scale perturbation into a base flow in which there was no natural particle mobilization, i.e., any subsequent particle mobilization was primarily driven by the introduced flow scales.

 \subsection{Canonical Flow}

  \begin{table*}[h]
    \centering
    \caption{A summary of the zero-pressure-gradient boundary-layer properties, i.e., the canonical flow}
    \begin{tabularx}{0.9\textwidth}{ >{\centering\arraybackslash}X | >{\centering\arraybackslash}X | >{\centering\arraybackslash}X | >{\centering\arraybackslash}X | >{\centering\arraybackslash}X |>{\centering\arraybackslash}X |>{\centering\arraybackslash}X |>{\centering\arraybackslash}X |>{\centering\arraybackslash}X }
         $Re_{\tau}$ & $U_{\infty}$ & $U_{\tau}$ & $U_{\tau}/\nu$  &$\delta$ & $f_{s}$ & $f_{s}^{+}$ & $T_{s}$ & $T_{s}$${U_{\infty}}/{\delta}$   \\
         & (ms$^{-1}$) & (ms$^{-1}$) &(\textmu m) &(cm) &(Hz) &  & (s) &  \\
        \hline
         1510 & 8.87 & 0.333 & 46.01 &  6.94 & 20000 & 0.3619 & 120 & 15320 \\
         \hline
    \end{tabularx}
    \label{tab:BLCharacteristics}
 \end{table*}
 
 To establish a baseline, the canonical smooth-wall flow is first characterized using hot-wire anemometry (HWA). All hot-wire measurements conducted as part of this work utilized a single wire probe with a diameter of 5~\textmu m. In keeping with the recommendations of \cite{Hutchins2009a}, the etched length of the hot-wire measured approximately 1.125~mm, thus yielding an aspect ratio of nearly 225. This single-wire probe operated in conjunction with a constant temperature anemometer system at an overheat ratio of 1.8. The signals were acquired using a digital data acquisition system at a sampling frequency of 20~kHz. To establish the precise wall-normal location, the traverse was fitted with a linear magnetic encoder, having a minimum resolution of 1~\textmu m. 
 
 At the streamwise location under consideration, the unforced (or unperturbed) flow over the smooth wall had a friction Reynolds number of $Re_\tau=~U_\tau \delta/\nu \approx~1510$. This determination relied on the boundary-layer thickness $\delta$ and friction velocity $U_{\tau}$ obtained by fitting the mean velocity profile to the composite velocity profile of \citet{Chauhan2009a}. Additional boundary-layer characteristics of this canonical wall-bounded turbulent boundary layer are summarized in Table~\ref{tab:BLCharacteristics}. For the rest of this work, this baseline flow field is referred to as the canonical flow.

 \subsection{Oscillating Airfoil Mechanism}
 
 Following this initial baseline characterization, the single-phase flow with the inclusion of the airfoil was considered. A NACA-0010 wing section (airfoil) with a chord length of 0.20~m was positioned at a height of approximately 0.15~m from the tunnel floor at the entrance to the test section. The leading edge of the wing coincided with the approximate end of the boundary-layer-trip. The airfoil underwent oscillations around a mean angle of attack of $A_0=-8^{\circ}$ which introduced large-scale structures into the flow. The spanwise axis about which the airfoil oscillated passed through the quarter-chord of the airfoil. 

  \begin{figure}[t]
     \centering
     \includegraphics[width=0.475\textwidth]{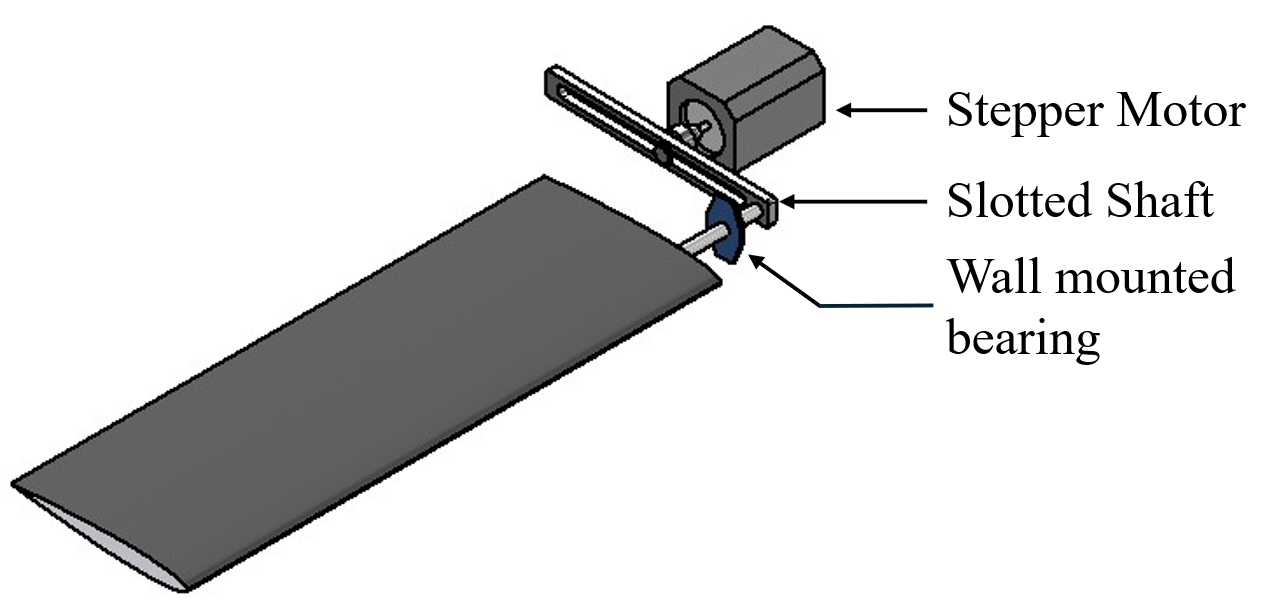}
     \caption{A schematic of the scotch-yoke mechanism used to oscillate the airfoil}
     \label{fig:OscillatingMechanism}
 \end{figure}

 The oscillatory mechanism employed was of the scotch-yoke type (see Fig.~\ref{fig:OscillatingMechanism}), where the rotary motion of a computer-controlled stepper motor (shaft) drove the reciprocating motion (oscillation) of the airfoil. Wall-mounted bearings, attached to the side walls of the wind tunnel, supported the airfoil while allowing it to swivel about the quarter-chord. A slotted shaft coupled the stepper motor with the airfoil, as shown in Fig.~\ref{fig:OscillatingMechanism}. The offset of the stepper motor shaft within the slotted shaft controlled the amplitude of oscillation $A_f$ of the airfoil, which in turn controlled the energy of the introduced flow-scale. The frequency of oscillation $f_f$ of the airfoil, and hence the introduced scale size, was regulated by the RPM of the stepper motor. Consequently, the oscillatory mechanism of the airfoil was engineered to facilitate independent regulation of both its frequency and amplitude, thereby enabling separate control over the wavelength and energy associated with the introduced scale. To measure the position of the airfoil as it traversed an oscillatory cycle, two light gates were employed to generate pulses when the airfoil reached either of its extremities of oscillation, i.e., the positions of maximum deflection. For the rest of this work, the flow field developing under the influence of this oscillating airfoil apparatus is referred to broadly as the forced flow. In particular, when the airfoil is present in the flow but not oscillating, i.e., $A_f=0$, the flow is referred to as non-periodic forced flow, recognizing that the wake of a still airfoil may still force the flow.

\begin{table*}[t]
    \centering
    \caption{A summary of boundary-layer properties with tonal free-stream forcing}
    \begin{tabularx}{\textwidth}{ >{\centering\arraybackslash}X |>{\centering}p{0.04\textwidth} | >{\centering\arraybackslash}X |>{\centering}p{0.045\textwidth} |>{\centering\arraybackslash}X | >{\centering\arraybackslash}X | >{\centering}p{0.045\textwidth} |>{\centering\arraybackslash}X |>{\centering\arraybackslash}X |>{\centering\arraybackslash}X |>{\centering\arraybackslash}X }
         $f_{f}$ & $A_{f}$ & $\delta$  & $\lambda_{f}$ & $U_{\infty}$ & $U_{\tau}$ & $U_{\tau}/\nu$  & $T_{s}$ & $T_{s}U_{\infty}/\delta$ & $f_{s}^{+}$ & $Re_{\tau}$  \\
        (Hz) & ($^{\circ}$) & (cm) & & (ms$^{-1}$) & (ms$^{-1}$) &(\textmu m) & (s) &  &  &\\
        
        \hline
        
         0 & 0 & 7.78 &        -     & 9.524 & 0.341 & 44.80 & 120 & 14869  & 0.381 & 1736 \\

        5 & 2 & 8.02 & 17.1$\delta$ & 9.095 & 0.343 & 44.93 & 240 & 27187  & 0.382 & 1787 \\
          
        10 & 1 & 7.77 & 8.7$\delta$ & 9.255 & 0.338 & 44.94 & 117.5 & 13990  & 0.376 & 1730 \\

                        
        10 & 2 & 7.65 & 9.0$\delta$ & 9.257 & 0.345 & 44.58 & 117.5 & 14209   & 0.382 & 1717 \\

        10 & 3 & 7.55 & 9.0$\delta$ & 9.105 & 0.342 & 44.99 & 117.5 & 14189   & 0.381 & 1676 \\

        15 & 2 & 8.33 & 5.4$\delta$ & 8.933 & 0.340 & 45.65 & 77.6  & 8326    & 0.371 & 1825 \\

        \hline

    \end{tabularx}
    
    \label{tab:HWAProfile}
\end{table*}

The forced single-phase flow was initially studied to establish its mean characteristics using hot-wire anemometry (HWA). Three oscillation frequencies - 5 Hz, 10 Hz, and 15 Hz - were examined. In addition, the effect of varying amplitude was investigated by testing three different amplitudes $A_f=1^{\circ},~2^{\circ}$ and $3^{\circ}$. These measurements of varying amplitude were conducted at a constant oscillation frequency of 10 Hz. The boundary layer characteristics at the measurement station for these test cases are summarized in Table~\ref{tab:HWAProfile}. Taylor's frozen hypothesis, with a convection velocity of $U_{c}=20U_{\tau}$~\citep{Hutchins2011a}, was used to convert the oscillation frequency $f_f$ into a forcing scale $\lambda_f$. Both $f_f$ and $\lambda_f$ are also summarized in Table \ref{tab:HWAProfile}. The friction velocity $U_{\tau}$ and the boundary-layer thickness $\delta$ were determined by fitting the mean velocity profile to a composite profile as described in \citet{Rodriguez2015a} and \citet{Esteban2017a}. This composite profile employed a wake formulation specific to flows with free-stream turbulence, originally proposed by \citet{Hancock1989a} and subsequently used in the friction velocity estimation of \citet{Esteban2017a}. The sampling duration was systematically adjusted, as detailed in Table~\ref{tab:HWAProfile}, in accordance with the oscillation frequency $f_f$ to guarantee that an equivalent number of oscillations were captured within each sampling interval. This approach ensured that an identical number of oscillating cycles were sampled while calculating the spectra. 

\subsection{Capturing Particle Mobilization}
 Following the characterization of the single-phase flow, with and without forcing, the attention was turned towards the dual-phase flow. Here, the term dual-phase denotes the flow over a particle bed. To study the initiation of particle mobilization, a particle bed (0.30~m wide $\times$ 0.127~m long $\times$ 0.009~m deep) was placed on the bottom wall of the test section. The leading edge of the particle bed was approximately 2.73~m downstream of the boundary-layer-trip (see Fig. \ref{fig:sch}). The particles were nominally spherical (85\%) grade IV soda-lime particles of density $\rho_{p}=1300\,$ kg m$^{-3}$ and manufacturer-specified diameter $d_p=$ 300 -- 425~\textmu m (\cite{MoSci2024}), with 80\% of the particles within the specified diameter range. 

Two sets of dual-phase measurements were conducted. The first set of measurements focused solely on the motion of particles on the surface of the particle bed. The second set of measurements captured the motion of both the particle-phase and carrier-phase simultaneously. A single 4~MP high-speed camera captured the motion of the particle-phase over both sets of measurements. The camera was placed over the roof of the tunnel and focused on the leading edge of the particle bed. A 200~mm (f/5.6) lens was used to focus on the particle bed. The resulting field of view (FOV) was $\Delta x \times \Delta y \approx $ 54.4~mm $\times$ 30.5~mm ($\Delta x \times \Delta y \approx 0.7\delta \times 0.4\delta$) at a pixel resolution of about 19.06 \textmu m px$^{-1}$.  The region of interest over the particle bed was illuminated using four high-intensity LEDs, while the camera operated at a frame rate of 400 Hz with an exposure period of 250~\textmu s. In this manner, 5477 images capturing the particle-phase (spanning a total sampling period of over 13.6 s) were acquired over each measurement set. 

Due to the variable nature of particle mobilization, repeated trials of each measurement set were performed. Before each measurement, the particle bed was replenished with soda-lime particles to compensate for any depletion from the previous set. The bed was manually leveled using the tunnel floor as a reference, and this was verified by ensuring that both the particle bed surface and the tunnel floor were in focus when viewed with the high-speed camera positioned over the tunnel roof. Care was taken not to alter the compactness of the particles, thereby ensuring that nominally identical conditions existed at the start of each measurement set.

Once images of the particle bed under the influence of the carrier-phase were obtained, the following procedure was employed to extract particle motion utilizing these raw images. First, individual images were registered to freeze the motion of the entire particle bed. This motion of the particle bed resulted from the relative motion between the tunnel wall and the camera because of tunnel vibration. Second, the registered images were analyzed to extract the motion of the particles using a standard Particle Image Velocimetry (PIV) algorithm applied between successive image pairs. To conduct this second step, the open-source PIV algorithm PIVLAB \citep{PIVLab2021a} was used. Beginning with an initial window size of 64\,$\times$ 64~px$^{2}$ and 50\% overlap between windows, three successive correlation passes were carried out. The interrogation window size was progressively reduced to a final interrogation window size of 16\,$\times$ 16~px$^{2}$ with 50\% overlap between windows. The nominally spherical particles spanned between 10 and 15 pixels in diameter. Therefore, the final interrogation window was large enough to encompass about 1.5 -- 2 particles. This step yielded an average displacement estimate over the interrogation area. 

To isolate large particle motions, the PIV-based displacement field was thresholded using a value of one pixel. This threshold value was approximately 8\% of the diameter of a single particle. This meant that no distinction was made between particles in different stages of particle mobilization. That is to say, any particle exhibiting oscillatory (rocking) or rotational (rolling) motion on or along the bed surface was classified as being mobilized, provided that its displacement exceeded the 1-pixel threshold. This procedure generated a displacement field for each time step, delineating areas on the particle bed that included particles with displacements exceeding the specified threshold. This displacement field was then organized into groups; particles that were adjacent to each other were considered part of a particle group or cluster. Next, clusters were further filtered based on the size of the clusters. Only clusters spanning more than three PIV interrogation windows in either the streamwise or spanwise direction were retained. The total number of these retained particle clusters across an image pair $N_g$ was measured and tracked, to demonstrate the efficacy of the current approach.

\subsection{Carrier-Phase Characterization}

A multi-camera PIV approach was used to measure the carrier-phase velocity field. A stack of two 4~MP high-speed cameras at a pixel resolution of 336 $\times$ 2560~px$^2$ each was utilized.  This ``long and skinny'' FOV yielded an array (or line) of instantaneous streamwise and wall-normal velocity vectors akin to a line of synchronous hot-wire measurements. For reference, this approach has been previously employed and described in detail in~\citet{Artham2021a}. This approach is referred to here as the line-PIV approach, as the end result of these measurements was instantaneous streamwise and wall-normal velocities along a single wall-normal line perpendicular to the wall.  

The lower camera, set with high magnification, employed a 200 mm (f/4) lens and focused on the near-wall region, achieving a FOV of $\Delta x \times \Delta z =$4.5~mm $\times$ 34~mm and a pixel resolution of 13.3~\textmu m $\mathrm{px}^{-1}$. In contrast, the higher camera, with lower magnification, used a 105 mm (f/4) lens, concentrating on the region further from the wall, with a FOV of $\Delta x \times \Delta z =$ 12~mm $\times$ 93~mm and a pixel resolution of 36.02~\textmu m $\mathrm{px}^{-1}$. Consequently, the combined FOV spanned approximately $\Delta x \times \Delta z =$ 12~mm $\times$ 115~mm ($\Delta x \times \Delta z \approx 0.15\delta \times 1.48 \delta$), highlighted by the red shaded area in Fig.~\ref{fig:sch}. A high-speed, dual-head Nd:YLF laser sheet of wavelength 527 nm illuminated this FOV. The laser operated at a 2 kHz pulse frequency, allowing a straddling time of 25~\textmu s between pulses. The cameras simultaneously operated at a frame rate of 2~kHz ($f_{s}^{+}\approx 3.805$) to acquire 29,600 image pairs over 14.8 seconds. The carrier-phase was seeded with water-based smoke aerosols, approximately 10 pixels in diameter.

The acquired particle images were processed using the solver DaVis 10 to obtain the time-dependent velocity fields. The procedure first involved background subtraction using a symmetric running average of 15 images. This was then followed by applying a multi-pass vector calculation algorithm. The initial window size was 64$\times$64~px$^{2}$ and the final window size was 24$\times$24~px$^2$ with a 50\% overlap. In terms of viscous scales, the final interrogation window sizes were $\Delta x^{+} \times \Delta z^{+} \approx$ 7 $\times$ 7 for the bottom camera and $\Delta x^{+} \times \Delta z^{+} \approx$ 19$\times$ 19 for the top camera. Here, the viscous scales used for normalization were those corresponding to the canonical flow (see Table~\ref{tab:BLCharacteristics}). At the end of these PIV processing steps, the velocity field (streamwise and wall-normal) was interpolated onto a single wall-normal profile that spanned the entire wall-normal extent of the boundary layer.

\subsection{Airfoil Oscillation Profiles}

In order to demonstrate the capabilities of the experimental framework, three different airfoil oscillation profiles were considered. Profile A involved continuous oscillations, while Profiles B and C comprised intermittent motions. In Profiles A and B, the amplitude was fixed at $A_f=2^\circ$; however, during Profile C, multiple amplitudes of oscillation $A_f$ were considered. During Profile A, the airfoil was subjected to a continuously increasing frequency of oscillation. Starting from rest, the oscillation frequency $f_f$ was steadily increased over 10 s to a terminal frequency before the oscillation was stopped. Two different terminal frequencies $f_f=10$ and 15 Hz were considered. 

Note, as previously stated, the airfoil oscillated around a mean position of $A_0=-8^\circ$ with respect to the free-stream flow. Both profiles B and C started with the airfoil in the extreme pitch-up position, i.e., the airfoil was positioned at $A_0+A_f$. During Profile B, the airfoil was rapidly pitched down to its extreme pitch-down position, i.e., $A_0-A_f$, and held in this position for varying periods of time. Subsequently, the airfoil was brought back to its original position $A_0+A_f$ and held stationary once again. This cycle continued for a period of three pitch-down and subsequent pitch-up motions, before the particle bed was replenished and re-leveled. During Profile B, the amplitude of oscillation was maintained constant at $A_f=2^\circ$.

Profile C also started with the airfoil in the extreme pitch-up position, followed by an abrupt motion to the extreme pitch-down position, i.e., from $A_0+A_f$ to $A_0-A_f$. However, in this case, the amplitude of oscillation varied from $A_f=1^\circ$ to $2.5^\circ$ in increments of $0.5^\circ$. All of the profiles of airfoil oscillation are summarized in Table~\ref{tab:OscProf}.

 \begin{table}[htbp]
     \centering
     \caption{Airfoil oscillation profiles considered}
     \begin{tabularx}{0.4\textwidth}{c|c|c|c}
         Profile & Type & $f_{f}$ (Hz) & $A_{f}$ \\
         \hline 
        A & Continuous & 0 -- 10 & $2^{\circ}$\\
        & & 0 -- 15  & \\
        
        B  & Intermittent & -- & $2^{\circ}$ \\
        
        C  & Intermittent & -- & $1^{\circ}$, $1.5^{\circ}$ \\
        & & & $2^{\circ}$, $2.5^{\circ}$ \\
        \hline
     \end{tabularx}
     \label{tab:OscProf}
 \end{table}

\section{Results}
\label{sec:results}
The results section is organized as follows. First, a detailed characterization of the single-phase (carrier-phase) flow is presented. This canonical flow served as the baseline for comparison. In particular, the structural changes caused are carefully characterized. Following this, the dual-phase flow is considered. Here, dual-phase refers to the flow field when the carrier-phase flow developed over a particle bed, irrespective of whether the particles are mobilized.

\subsection{Characterization of the Forced Carrier-Phase}

\subsubsection{Mean Statistics}

The forced carrier-phase flow was characterized using HWA to clearly establish the changes in the flow resulting from the oscillation of the airfoil. The friction velocity $U_\tau$ as the airfoil oscillated at various frequencies is summarized in Table~\ref{tab:HWAProfile} while the corresponding canonical smooth-wall friction velocity is tabulated in Table~\ref{tab:BLCharacteristics}. The maximum variation in the friction velocity with respect to the canonical flow was $3.60\%$ (when $A_f=2^\circ$, $\lambda_f\approx 9.02 \delta$). This variation is of the order of the measurement error for skin-friction measurements \citep{Naughton2002a,Esteban2017a}. Hence, it was concluded that all the friction velocities in the present work lay within the experimental uncertainty, and no further inference was drawn. 

\begin{figure}[htpb]
\centering
\includegraphics{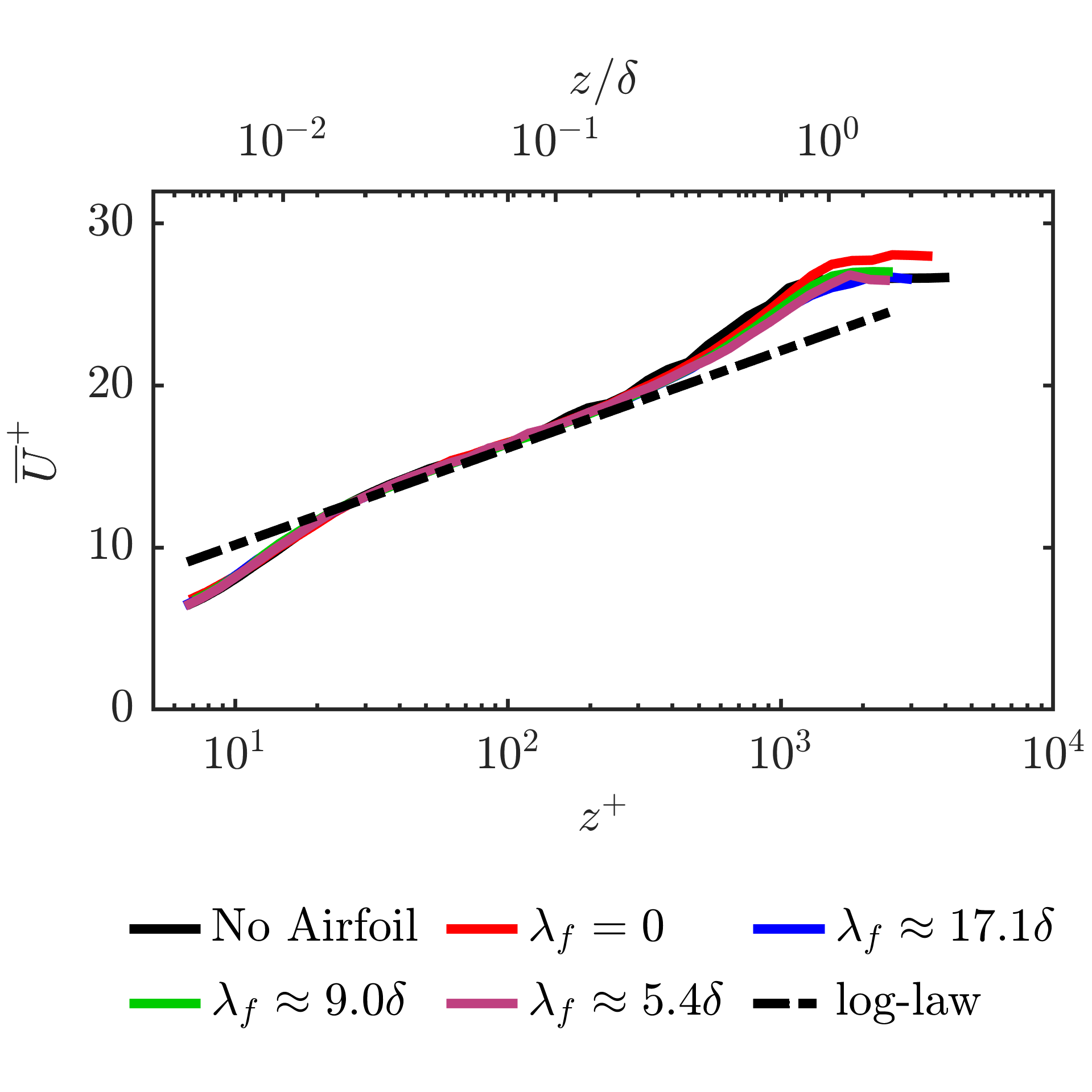}
\caption{Comparison of the mean velocity profiles with and without free-stream forcing ($A_{f}=2^{\circ}$). The black dotted line indicates the log-law, $U^{+}=1/\kappa\ln{z^{+}}+A$, with coefficients $\kappa=0.384$ and $A=4.17$ \citep{Chauhan2009a} }
\label{fig:VelProfileComparison}
\end{figure}

The mean velocity profiles in inner scaling are shown in Fig. \ref{fig:VelProfileComparison} when the amplitude of oscillation was $A_{f}=2^{\circ}$. The profiles corresponding to the three different forcing wavelengths considered are shown along with the canonical and non-periodic forced flow. A log-law profile with coefficients $\kappa=0.384$ and $A=4.17$ \citep{Chauhan2009a} is also shown for reference. All the profiles collapsed well in the near-wall and the logarithmic region. Further away from the wall, in the wake region, some differences were observed between the canonical flow and the forced flows.

\begin{figure}[htbp]
\centering
\includegraphics{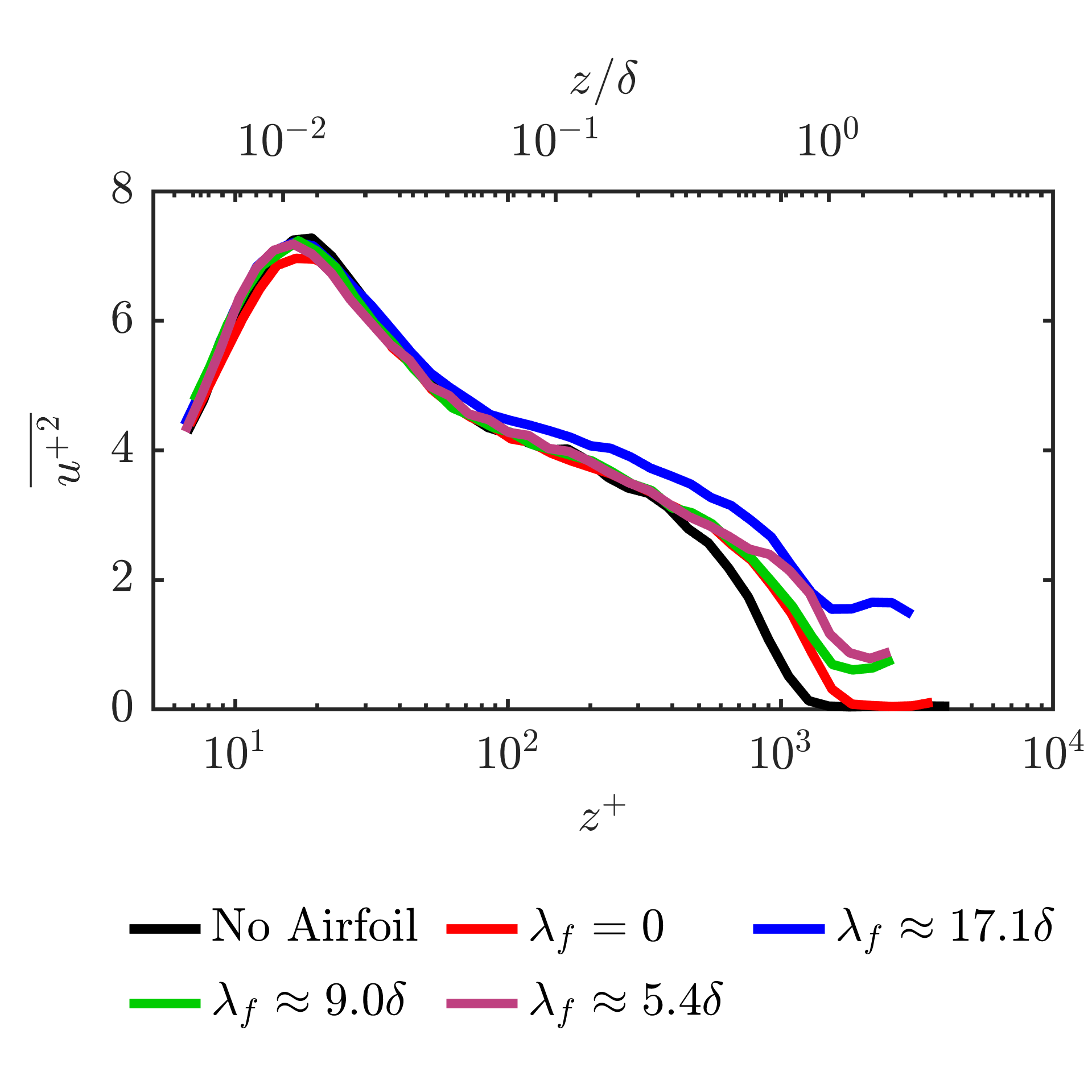}
\caption{Comparison of the turbulence intensity profiles of the canonical and forced flow ($A_{f}=2^{\circ}$)}
\label{fig:TIComparison}
\end{figure}

The corresponding streamwise broadband turbulence intensity profiles in inner scaling are shown in Fig. \ref{fig:TIComparison}. All profiles collapsed reasonably well in the near-wall region, with no clear changes in the inner peak. However, moving away from the wall into the logarithmic region and beyond, the profiles started to deviate from the canonical flow. Clear differences were observed in the wake region, indicative of the enhanced free-stream turbulence resulting from the oscillation of the airfoil. In particular, the canonical and non-periodic forced flow ($A_f=0^\circ$) are observed to have very low free-stream turbulence ($<1\%$). All other forced flows showed an increase in the free-stream turbulence. The largest increase in free-stream turbulence was observed in the case of the flow with the longest forcing wavelength ($f_f=5$~Hz, $\lambda_f\approx 17.1 \delta$).

\subsubsection{Spectra}

\begin{figure}[htbp]
\centering
\includegraphics{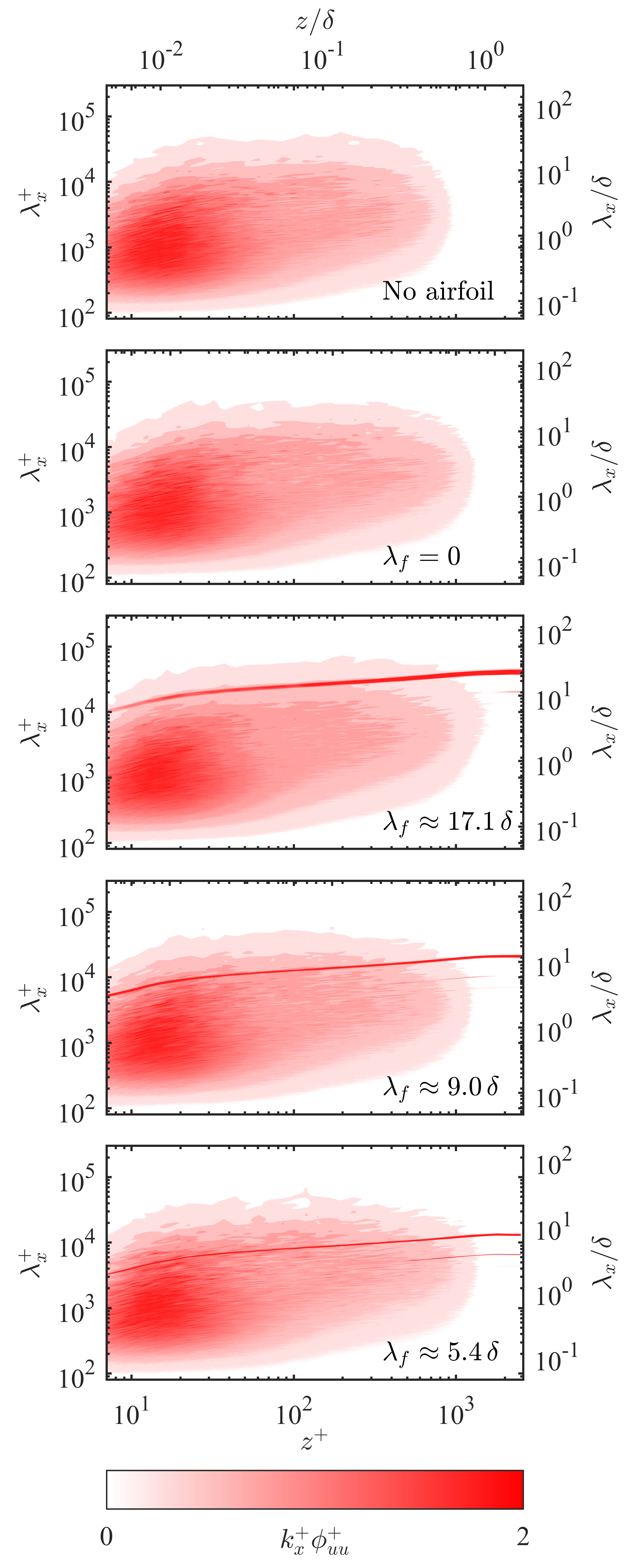}
\caption{Contours of one-dimensional pre-multiplied energy ($k_{x}^{+}\phi_{uu}^{+}$) when $A_{f}=2^{\circ}$. Spectra corresponding to the canonical and non-periodic forced flow are also presented for reference}
\label{fig:SpectraComparison}
\end{figure}

The streamwise fluctuations are decomposed into their flow-scale components to understand the changes observed in the turbulence intensity profiles. The velocity fluctuations from the hot-wire measurements are first decomposed into their Fourier frequency components. Following this, Taylor's frozen turbulence hypothesis is used to convert these frequencies into constituent wavelengths $\lambda_{x}=2\pi/k_{x}=2\pi \overline{U}/f$. Here, $\overline{U}$ is the local mean velocity at a given wall-normal location, $k_x$ is the streamwise wavenumber, and $f$ is the corresponding flow frequency. The contours of the viscous-scaled one-dimensional energy in pre-multiplied form $k_{x}^{+}\phi_{uu}^{+}$, are presented as a function of the wall-normal distance and the streamwise wavelength in Fig. \ref{fig:SpectraComparison}. It is again noted that the results presented pertain to a constant oscillation amplitude of $A_f=2^{\circ}$.

The overall structure of the boundary layer, as indicated by the spectra in Fig. \ref{fig:SpectraComparison}, remained largely unchanged with the introduction of the forcing. The presence of a distinct inner peak (centered at $\lambda_{x}^{+} \approx 1000$, $z^{+}\approx15$) for all cases is clearly observed. The location of this peak corresponded to the location of the near-wall peak in the broadband intensity curves (Fig. \ref{fig:TIComparison}). The spectral signature associated with this near-wall cycle did not change in any significant sense despite the presence of the airfoil. Hence, the near-wall turbulence production dynamics is not modified in a significant manner. A closer look at the spectra showed that the spectral signature of the canonical flow and the non-periodic forced flow were nearly identical. In contrast, all of the forced flows displayed a clear signature of the forcing scale in the form of excess energy in frequency bands associated with the oscillation frequency.

\begin{figure}[htpb]
\centering
\includegraphics{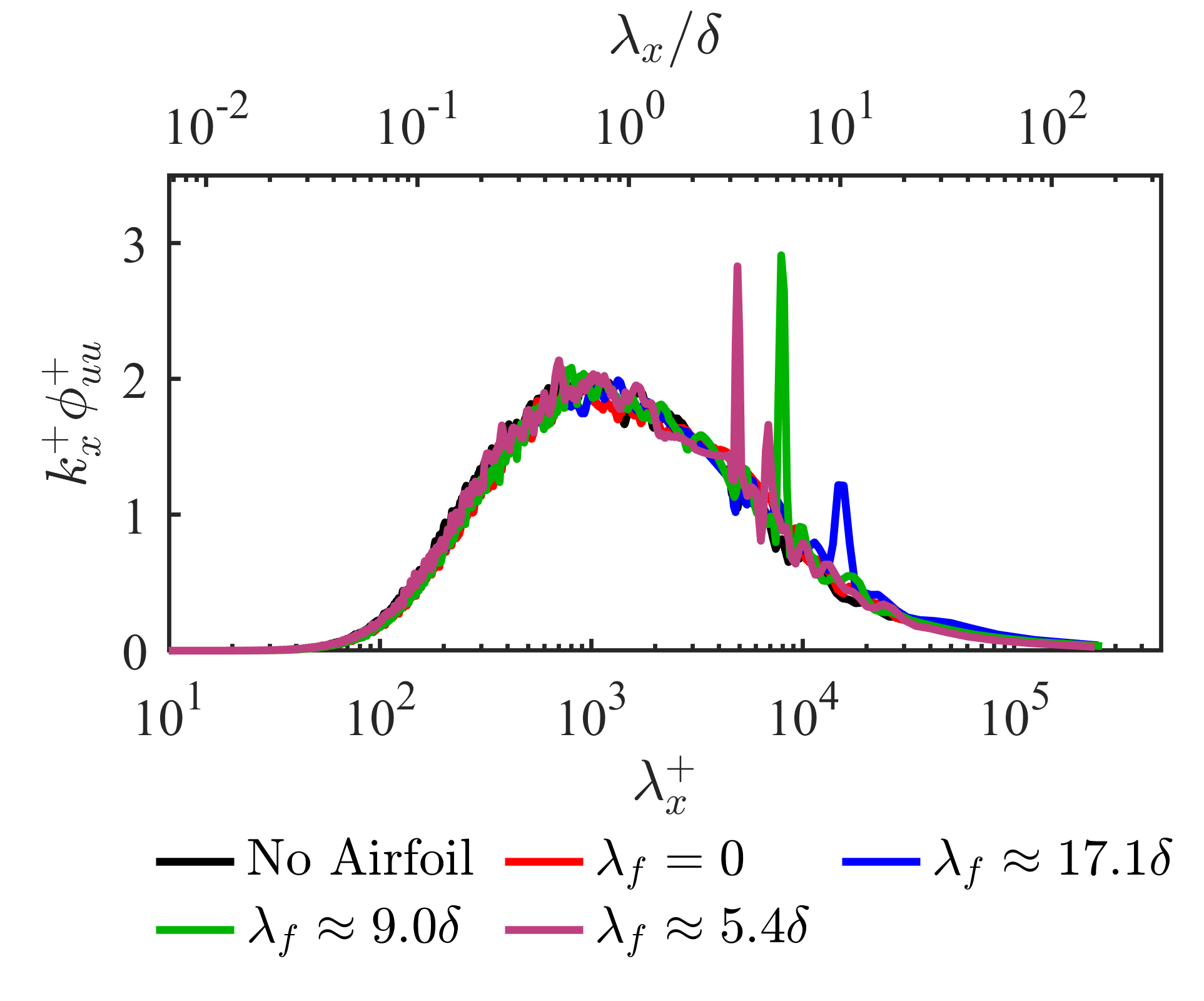}
\caption{Comparison of the energy spectra at $z^{+}\approx15$ across all cases with $A_{f}=2^{\circ}$. The energy spectra for the non-periodic forced flow and the canonical flow is also shown }
\label{fig:LinePlotSpectraComparison}
\end{figure}

The frequency band of enhanced energy represented the linear response of the boundary layer to the forcing~\citep{Duvvuri2015a,Bhatt2020a}. This band of excess energy extended from the free-stream and penetrated into the boundary layer, reaching the wall. This finding was consistent with studies that focused on broadband free-stream forcing, where the forcing scales were observed to reach the wall~\citep{Sharp2009a,Dogan2016a}. This aspect is also highlighted in Fig.~\ref{fig:LinePlotSpectraComparison} where the pre-multiplied spectra at $z^{+}\approx15$ are plotted to illustrate the near-wall spectral signature of the forcing scale. Note that in Fig.~\ref{fig:LinePlotSpectraComparison} the profiles have been smoothed using a Savitzky-Golay filter (spanning $\pm 4.2$ Hz) except around the forcing scales (and their harmonics) to underscore their signature. 
It is also observed that, apart from the excess energy at the forcing frequencies, the spectral content of the flows is nearly identical across the entire wavelength space. Only in the case of the smallest forcing wavelength ($f_f=15$~Hz, $\lambda_f\approx 5.4 \delta$) did the non-linearities develop to form a higher harmonic. This reinforces the interpretation that the forcing does not significantly alter the near-wall cycle.

\begin{figure}[htpb]
\centering
\includegraphics{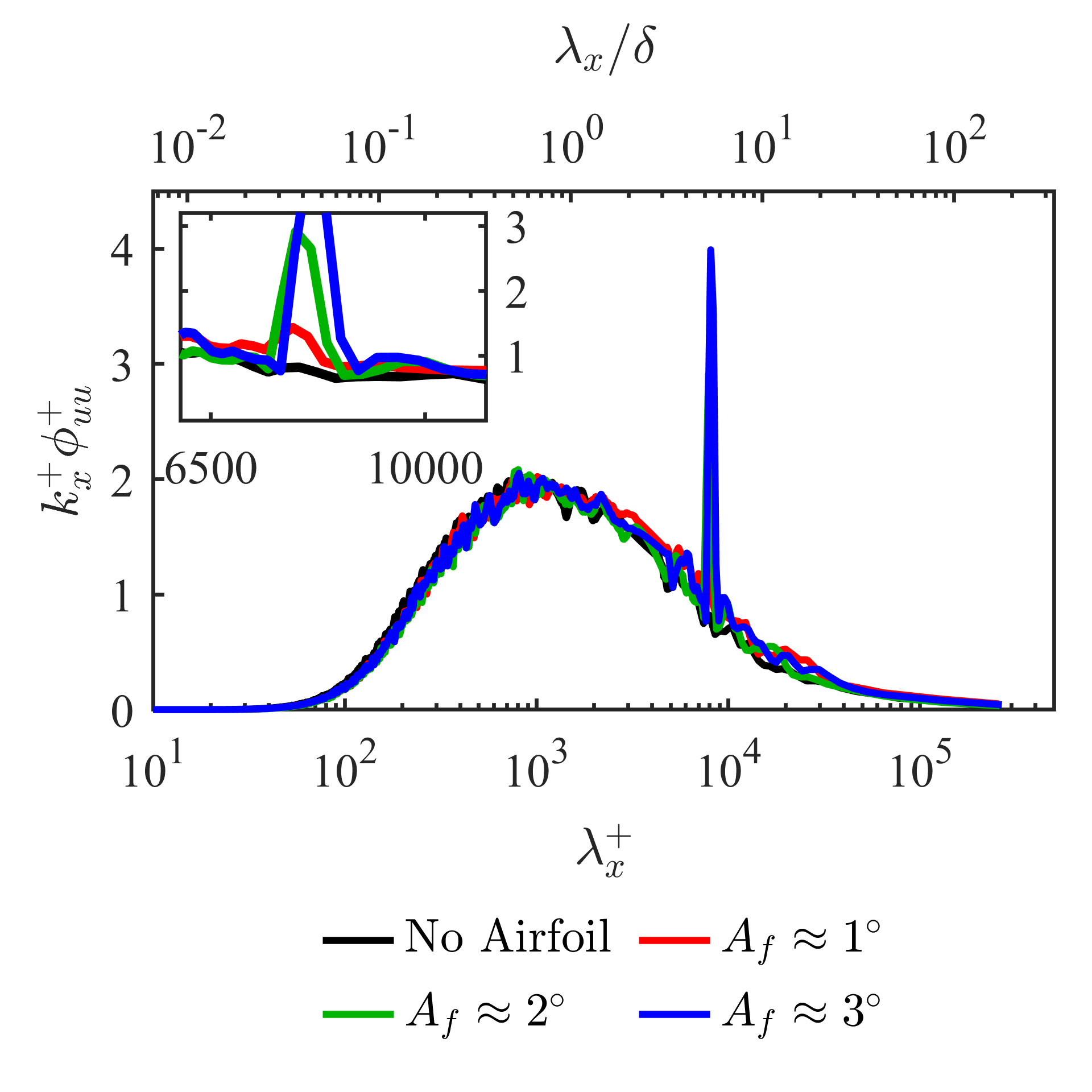}
\caption{Comparison of the energy spectra at $z^{+}\approx15$ when the oscillation amplitude was varied ($\lambda_{f}\approx9.02\delta$). The spectra for canonical flow is also shown for reference}
\label{fig:LinePlotSpectraComparison_Amplitude}
\end{figure}

To understand the relationship between the energy of the forcing scale and particle mobilization, it is desirable to vary the energy of the forcing scale. For this purpose, three different oscillation amplitudes, $A_f \approx 1^\circ$, $2^\circ$, and $3^\circ$, were considered, while the forcing frequency, $f_f$, and the corresponding forcing wavelength were kept constant at $\lambda_f \approx 9.02\delta$. Figure \ref{fig:LinePlotSpectraComparison_Amplitude} presents profiles of the pre-multiplied energy spectra measured at a wall-normal position of $z^+ \approx 15$ in the case of each of the amplitudes $A_f$ considered. The spectrogram corresponding to the canonical flow is also presented for reference. It is clear that, except for the distinct peak at the forcing scale, the energy distribution across the entire wavelength space is virtually identical when compared with the canonical flow, irrespective of the forcing frequency. At the forcing frequency itself, a distinct peak is observed (see also inset in Fig \ref{fig:LinePlotSpectraComparison_Amplitude}). The magnitude of this peak varied with the forcing amplitude $A_f$, where the largest magnitude corresponded with the largest forcing amplitude and vice versa. Hence, the oscillating airfoil mechanism was able to introduce flow scales of controllable wavelength and energy into the boundary layer, and its footprint was observed at the wall.

\subsubsection{Isolated Airfoil Motion}

\begin{figure*}[htbp]
\centering
\includegraphics{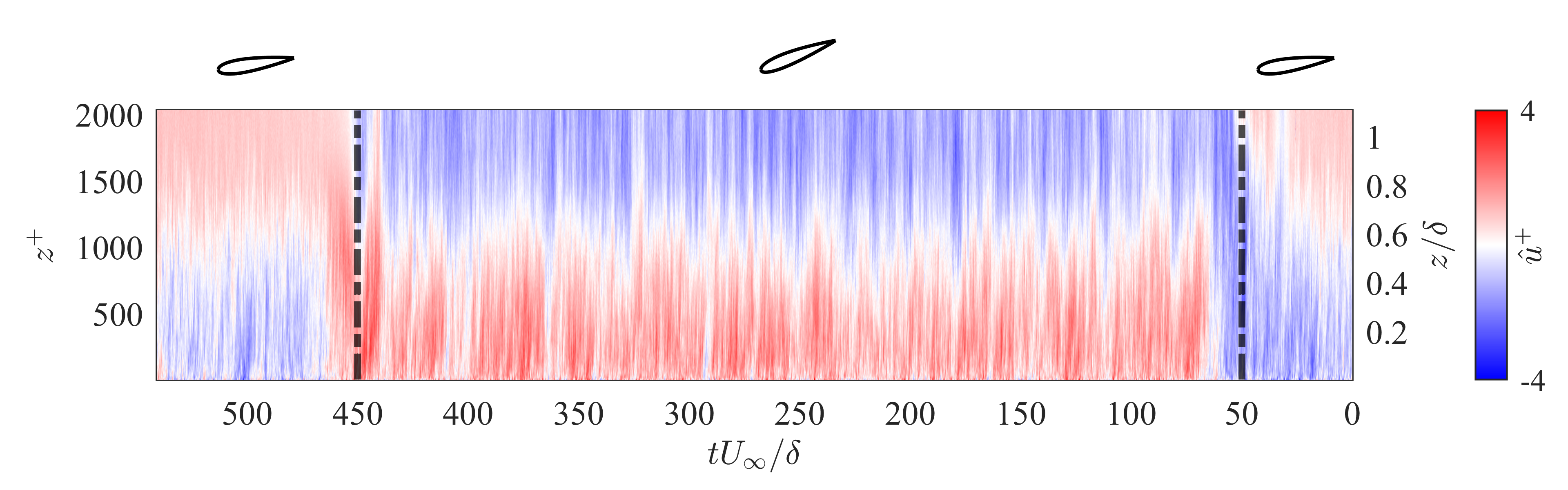}
\caption{Streamwise ensemble averaged velocity fluctuations $\hat{u}$ in the case of isolated motions of the airfoil (see Fig.~\ref{fig:PIVSinglePitchUp_wVel} for $\hat{u}$). The dotted black lines indicate the approximate instant of the pitch-down (at $tU_{\infty}/\delta\approx$~50) and  the pitch-up (at $tU_{\infty}/\delta\approx$~450) motion of the airfoil. The direction of the abscissa has been flipped to maintain a left-to-right flow perspective}
\label{fig:PIVSinglePitchUp}
\end{figure*}

\begin{figure*}[htbp]
\centering
\includegraphics{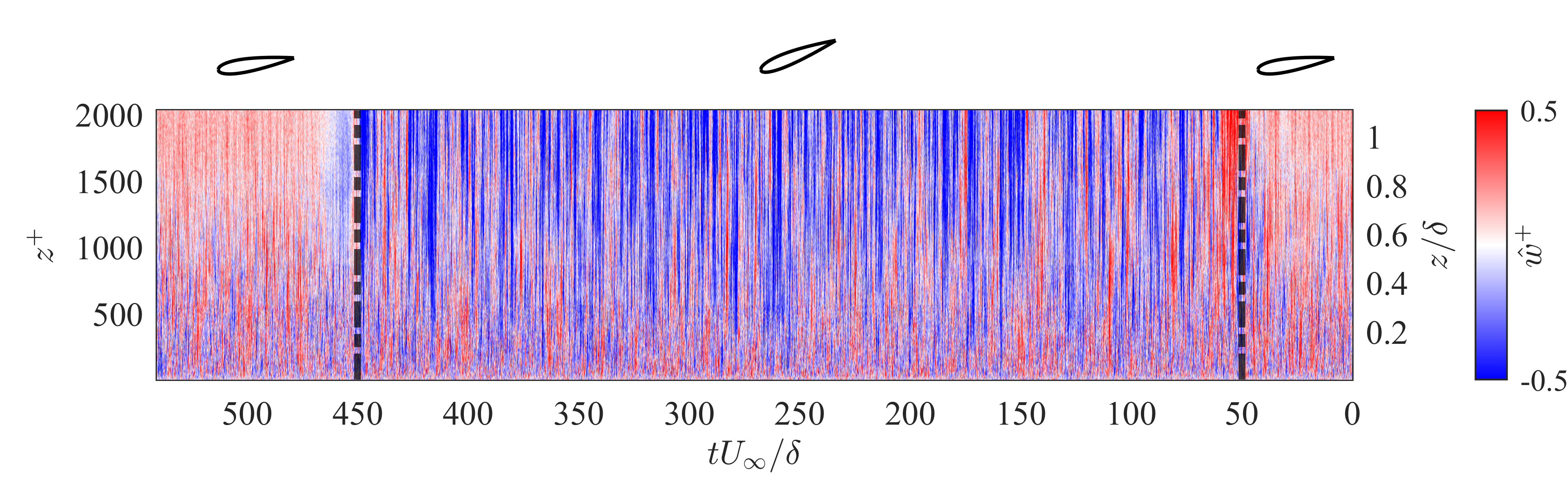}
\caption{Wall-normal ensemble averaged velocity fluctuations $\hat{w}$ during isolated motions of the airfoil (see Fig.~\ref{fig:SinglePitchUp} for $\hat{u}$). The direction of the abscissa has been flipped to maintain a left-to-right flow perspective}
\label{fig:PIVSinglePitchUp_wVel}
\end{figure*}
 
The HWA-based measurements presented so far investigated the structure of the carrier-phase when subjected to continuous periodic forcing. The PIV-based measurements offered an opportunity to characterize the flow from a different perspective. In particular, changes in the flow were investigated when the airfoil was subjected to a single impulsive pitch-down motion followed by a period of rest and a pitch-up motion (Profile B in Table~\ref{tab:OscProf}). This motion was chosen because it was observed that the particles appeared to be mobilized predominantly during the pitch-down phase of the motion. The airfoil was initially placed in the extreme pitch-up position and then impulsively pitched down. The airfoil was then held in the pitch-down position for about 400 eddy-turnover time scales (approximately 3\,s) and subsequently moved back into the pitch-up position. As the airfoil underwent these series of motions, the line-PIV approach was used to characterize the entire wall-normal extent of the flow (see schematic in Fig.~\ref{fig:sch} for FOV). Seven trials were conducted to obtain an ensemble average of the changes in the flow structure. 

Fig. \ref{fig:PIVSinglePitchUp} shows the mean subtracted streamwise ensemble average $\hat{u}^{+}$, as a function of time, over the entire wall-normal extent of the boundary layer. Note that the direction of the abscissa has been reversed to maintain a left-to-right flow perspective. The dotted lines indicate the upstroke of the airfoil followed by the downstroke of the airfoil, respectively. It is evident that the downstroke of the airfoil generated a low-speed region (relative to the local mean) in the free-stream and outer wake regions. However, for most of the wall-normal extent of the boundary layer, the airfoil pitch-down position is associated with a high-speed region (relative to the local mean). In contrast, when the airfoil is in the pitch-up position, the opposite is true, i.e., there is a high-speed region in the free-stream associated with a low-speed region within the boundary layer. As the airfoil transitions from its pitch-down to its pitch-up position, the evolving flow state is also captured. Most notably, it appears that when the airfoil moves from its pitch-up position to its pitch-down position, the low-speed flow in the boundary layer transitions out into the free-stream, i.e., there is an interruption of the high-speed region in the free-stream. The converse is true when the airfoil transitions from the pitch-down to the pitch-up state, i.e. the low-speed region in the free-stream gets interrupted. 

The corresponding ensemble average of the wall-normal velocity fluctuations (about the mean) $\hat{w}$, as a function of the entire wall-normal extent of the boundary layer, is presented in Fig. \ref{fig:PIVSinglePitchUp_wVel}. Similar to the behavior of the streamwise velocity fluctuations, it is observed that the outer wake of the boundary layer transitioned from a period of positive wall-normal fluctuations to a period of negative fluctuations following a pitch-down motion of the airfoil. In other words, when the airfoil is in its pitch-down position, there is a flow (albeit a weak one) towards the wall. However, unlike the streamwise velocity fluctuations, no clear organization in the wall-normal fluctuations was observed in the near-wall region. Hence, it appears that the near-wall increase in the streamwise velocity is a result of this weak flow of high-momentum fluid from the free-stream towards the wall when the airfoil is pitched down. 

\begin{figure*}[htpb]
\centering
\includegraphics{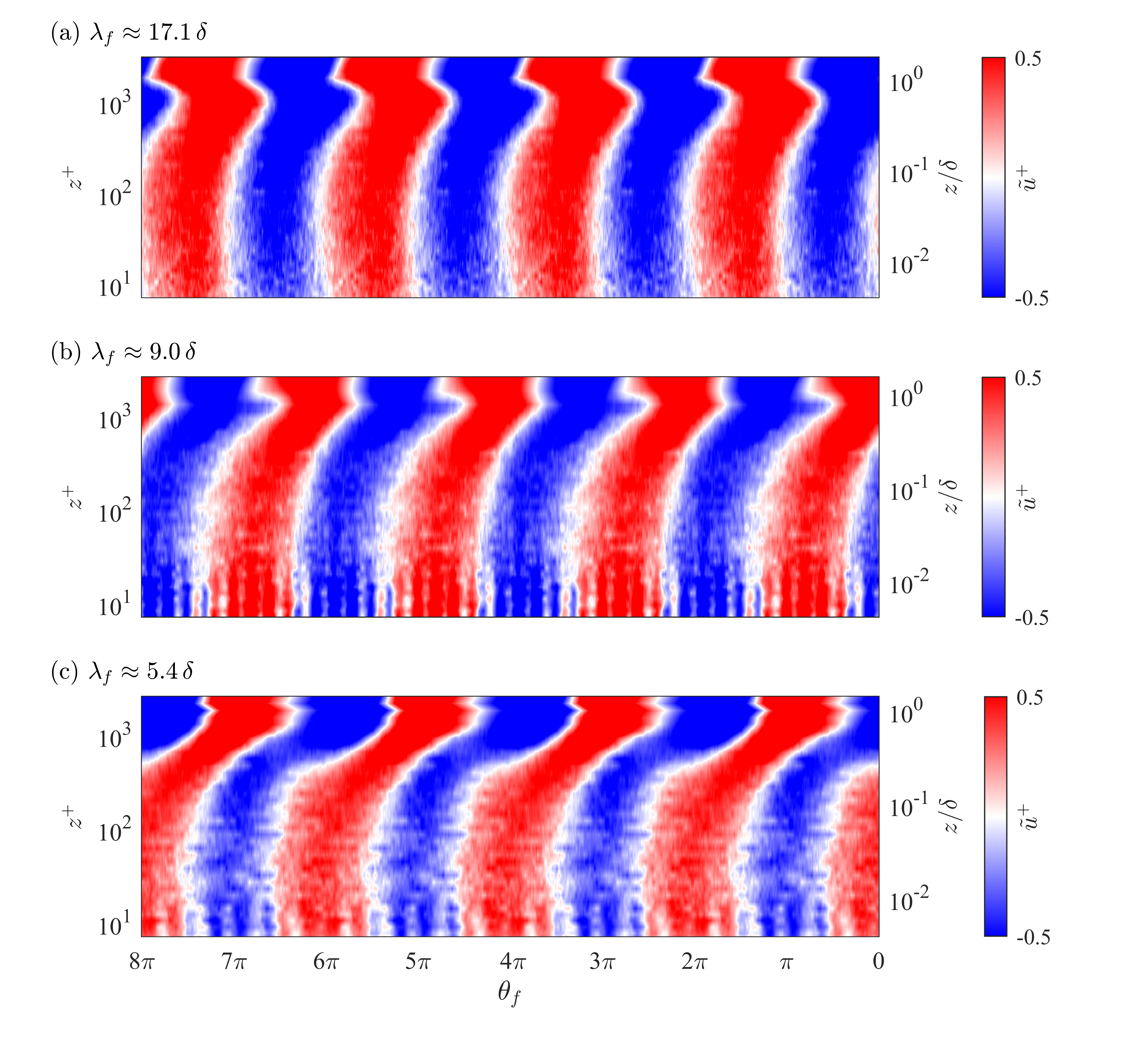}
\caption{ Phase-locked streamwise velocity fluctuations $\tilde{u}$ in the case of varying forcing wavelengths: (a) $\lambda_{f}\approx$~17.1$\delta$, (b) $\lambda_{f}\approx$~9.02$\delta$, and (c) $\lambda_{f}\approx$~5.4$\delta$, over four oscillation cycles of the airfoil. The abscissa is flipped to maintain a left-to-right flow perspective}
\label{fig:PhaseLocked}
\end{figure*}

\begin{figure*}[htpb]
\centering
\includegraphics{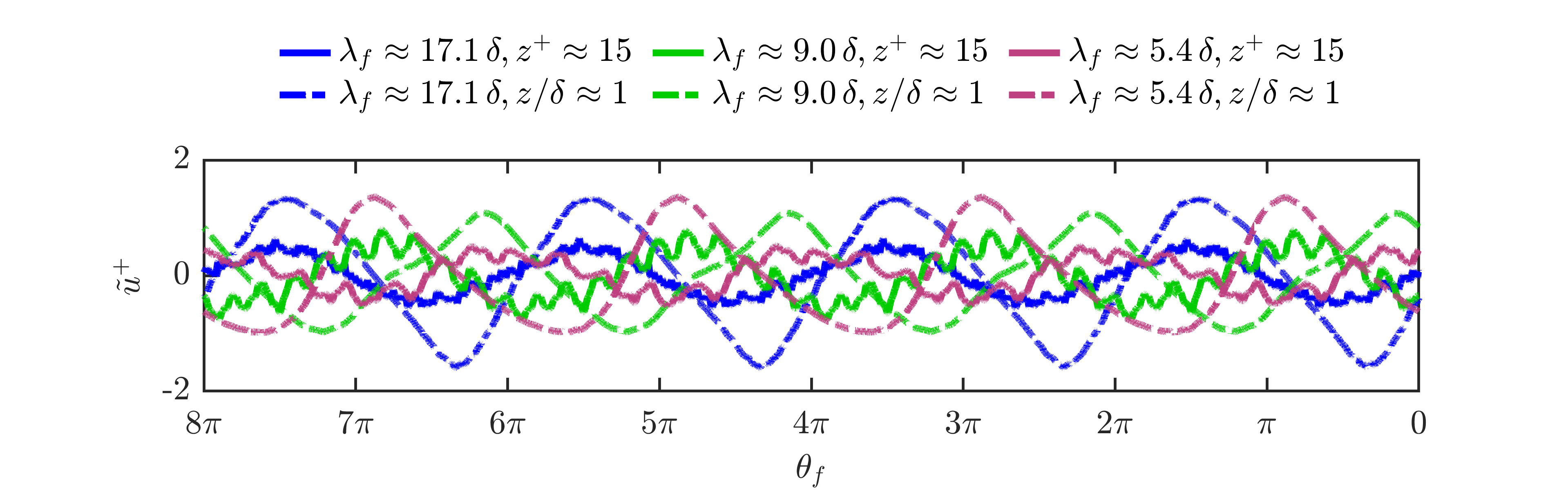}
\caption{ Phase-locked streamwise velocity fluctuations $\tilde{u}$ in the case of different forcing wavelengths are shown at two different wall-normal locations over four oscillations of the airfoil. The solid lines denotes $\title{u}$ at the location of near-wall peak ($z^{+}\approx15$) while the dotted line denotes that in the free-stream ($z\approx\delta$)}
\label{fig:PhaseLockedLinePlot}
\end{figure*}

\subsubsection{Phase-Locked Statistics}

Having established that the airfoil pitch-down state is associated with enhanced momentum in the wall region, the focus is turned once again to the scenario where the flow is being subjected to continuous periodic forcing by the airfoil and the associated HWA-based measurements. Since the external forcing is oscillatory in nature, following prior work (see for example~\citet{Reynolds1972a,Duvvuri2015a,Bhatt2020a}), the velocity field is triple-decomposed, i.e., into the mean velocity $\overline{U}$, the phase-locked fluctuating velocity component $\Tilde{u}$, and the broadband turbulent fluctuations $u$. The position of the airfoil (obtained from the infrared photo gate sensors) is used as the reference signal for the phase-locking procedure. The phase-locked components are then ensemble averaged over four oscillation cycles of the airfoil. Contours of these phase-locked velocity fluctuations corresponding to all three forcing cases studied are presented as a function of the wall-normal distance in Fig. \ref{fig:PhaseLocked}. Line profiles of the phase-locked velocity at $z^+\approx15$ (near-wall region) and $z\approx \delta$ (free-stream) are also shown in Fig.~\ref{fig:PhaseLockedLinePlot}. It is again noted that the direction of the abscissa has been reversed to maintain a left-to-right flow perspective. 

The contours of the phase-locked velocity fluctuations indicated that they comprised alternating periods of positive and negative velocity fluctuations for all the forcing wavelengths tested. Based on the aforementioned PIV measurements (refer to Fig.~\ref{fig:PIVSinglePitchUp}), it is concluded that every pitch-down motion of the airfoil is associated with a period of positive velocity fluctuations (high-momentum) within the boundary layer and vice versa.  More relevant to the current application is the fact that these zones of positive and negative fluctuations are observed to span the entire wall-normal extent of the flow, reaching the wall. However, the magnitude of fluctuations in the free-stream is higher than that near the wall (see Fig.~\ref{fig:PhaseLockedLinePlot}). While the persistence of the forcing scale across the entire wall-normal extent of the flows is clearly observed, differences are noted as the forcing wavelength is changed. These differences are more easily visualized in the phase-locked velocity profiles of Fig.~\ref{fig:PhaseLockedLinePlot}. First, in the near-wall region, the higher harmonics are more evident in the case of the shortest forcing wavelengths $\lambda_f\approx5.4\delta$ and $\lambda_f\approx 9.0 \delta$. This is consistent with the corresponding profiles of the spectra in Fig.~\ref{fig:LinePlotSpectraComparison} where the non-linearities are more evident in the near-wall region when $\lambda_f\approx 5.4\delta$. Interestingly, the phase-locked velocity corresponding to the longest wavelength $\lambda_f\approx9.0\delta$ had the largest amplitude in the free-stream. This points to the boundary layer showing differences in receptivity when the forcing scale is changed.

Several studies have focused on perturbations to boundary layers through various means. Some of the approaches introduced perturbations via the free-stream~\citep{Sharp2009a,Dogan2016a}, through surface oscillations~\citep{Jacobi2011a,Duvvuri2015a} or acoustic forcing~\citep{Bhatt2020a,Artham2021a}. In the present work, tonal forcing (i.e., forcing at a fixed scale) was introduced via the free-stream. Following prior work~\citep{Jacobi2011a, Duvvuri2015a,Bhatt2020a} the present observations are interpreted as follows. The boundary layer responds to the forcing linearly through an artificial or ``synthetic'' \citep{McKeon2018a} flow scale at the fundamental forcing frequency. These are the flow scales isolated by the phase-locking procedure, which are also identified by the peak in the spectra. As the flow then develops downstream, the non-linearities develop, leading to the observed harmonics. 

In the current application, this synthetic flow scale served as a means to introduce, into the boundary layer, a flow scale of a specific frequency (or wavelength). This introduced scale had excess energy at the forcing frequency $f_f$ when compared to the canonical flow. In addition, the level of this excess energy could be controlled by changing the amplitude of the oscillating airfoil $A_f$. Significantly, this synthetic scale penetrated the boundary layer, reaching the wall and generating shear at the wall, at a fixed controllable frequency. Further, the near-wall dynamics are not significantly modified by the introduction of the artificial scale. Hence, collectively, this forced flow provided a base flow with energetic flow structures of a given scale and energy that could then interact with a particle bed to mobilize particles. 

To provide more context, as the Reynolds number of canonical flows increased, the energy in the large-scales of the flow increased while the energy in the smaller scales of the near-wall cycle was not observed to increase~\citep{Smits2011a}. As a result, boundary layers in the field show considerably more energy in large-scale structures compared to laboratory flows. Figures ~\ref{fig:SpectraComparison}, \ref{fig:LinePlotSpectraComparison} and \ref{fig:LinePlotSpectraComparison_Amplitude} all demonstrate that the energy in the forcing scale is significantly higher than that present in the canonical laboratory boundary layer. Thus, the enhanced energy observed at the forcing scales allows for the ability to mimic particle mobilization observed at higher Reynolds numbers in a lower Reynolds number laboratory flow. In particular, energetic large-scales, which are the flow scales that mobilized large and heavy particles, can be considered. The following section presents demonstrative measurements of the same, where large and heavy particles are mobilized by the synthetic large-scales.

\subsection{Dual-Phase Measurements}

\begin{figure*}[htbp]
\centering
\includegraphics{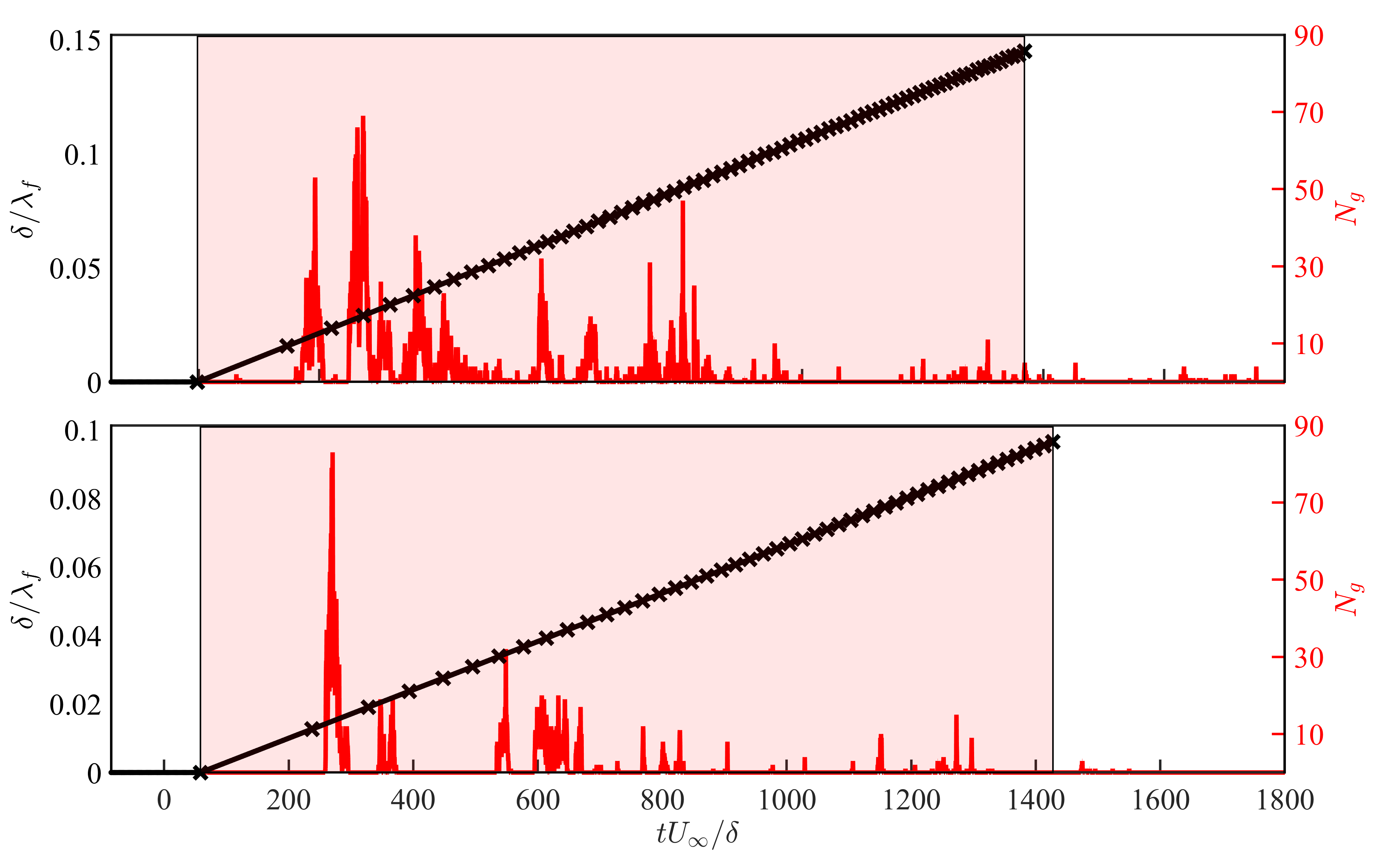}
\caption{Particle motion over oscillation Profile A, where the size of the introduced scale ($\lambda_{f}$) continuously decreases to a terminal scale-size of 5.4$\delta$ (top) and 9.0$\delta$ (bottom). The forcing length scale (\textbf{\textcolor{black}{---}}) is depicted on the left ordinate. The number of mobile particle clusters $N_g$ (\textbf{\textcolor{red}{---}}) is shown on the right ordinate. The pink-shaded region indicates the duration of the oscillation}
\label{fig:ContinousRampUp}
\end{figure*}

\subsubsection{Oscillation Profile A}
As a first step, the ability of the current framework to mobilize otherwise immobile particles is demonstrated. As previously described, the particle bed is filled with soda lime particles. The tunnel speed is carefully increased until a nominal free-stream velocity of about 9~ms$^{-1}$ is reached. This free-stream velocity was chosen, as there was very little to no particle mobilization at this velocity. In other words, at a nominal free-stream velocity of 9~ms$^{-1}$, the mean shear at the wall does not cross the static threshold needed to initiate particle mobilization. The tunnel was then run at this speed for a few minutes to allow all transients to die out. The oscillation frequency of the airfoil was then continuously increased from zero to a terminal frequency, followed by its abrupt stoppage (Profile A in Table~\ref{tab:OscProf}). Two terminal frequencies, 10~Hz and 15~Hz, were considered, as described in Table \ref{tab:OscProf}.

The number of mobile particle clusters $N_g$ as a function of time is shown in Fig.~\ref{fig:ContinousRampUp}. In addition, the oscillating frequency, presented as an inverse of the non-dimensional oscillation wavelength, is also shown. Note, the top plot had an oscillation frequency that spanned from 0 to 15 Hz ($5.4\delta$) while the bottom plot spanned frequencies from 0 to 10 Hz ($9.02\delta$). A common trend observed in both cases was that there is little to no mobilization in the white (unshaded) regions of the graph, i.e., when the airfoil is not in oscillation. Once the airfoil started oscillating, mobilization of particles is clearly observed. The mobilization activity is found to be intermittent. In both cases, it is also observed that there is a significant amount of particle mobilization during the initial stages of oscillation, followed by a gradual decrease. The introduced flow scales during this initial intense period of particle mobilization were of low frequency or large length scale. As the oscillation frequency progressed to large frequencies (longer wavelengths) fewer particles were mobilized. However, it is noted that in the initial phases of mobilization, particles whose placement on the particle bed is such that they are easy to mobilize, will be mobilized quickly. Here, the term easy is used to refer to particles that are held weakly by inter-particle forces purely on account of their random placement on the surface of the particle bed.  

\begin{figure*}[htpb]
\centering
\includegraphics{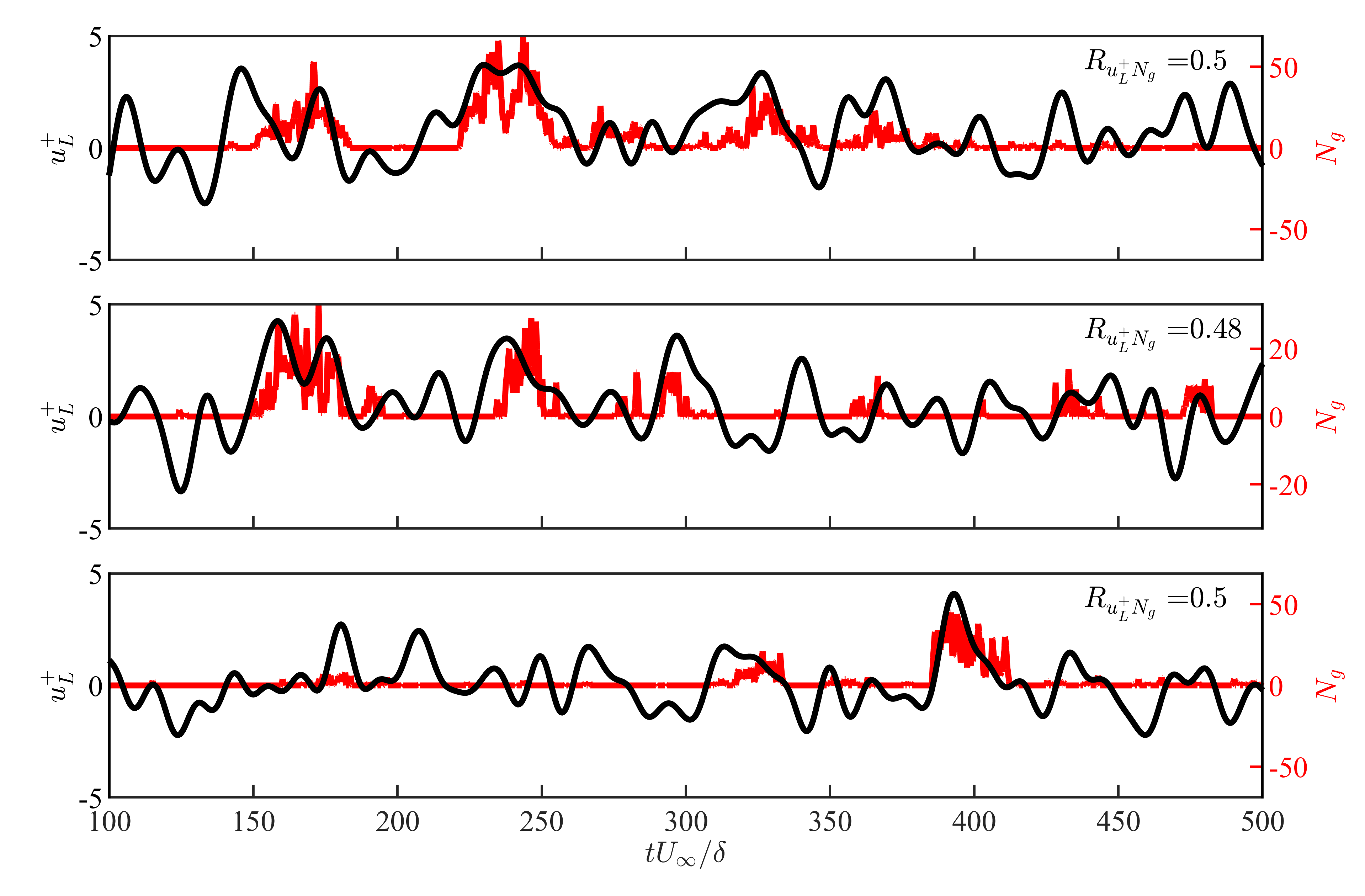}
\caption{Correlation between the number of mobile particle clusters $N_g$ and the extremely large-scale fluctuations $u_L$ of the carrier-phase. The number of mobile particle clusters $N_{g}$ is shown on the right ordinate (\textbf{\textcolor{red}{---}}). The extremely large-scale ($\lambda_{x}\geq$10$\delta$) fluctuations $u_{L}^{+}$  at $z^{+}\approx15$ is shown on the left ordinate (\textbf{\textcolor{black}{---}}). The abscissa denotes the flow-time in terms of eddy turnover time scales. Each subplot represents a different realization of the same experiment}
\label{fig:FluctuationPIV}
\end{figure*}

A combined PIV particle mobilization study investigated further the specific role that the carrier-phase eddies played in particle mobilization. As the airfoil underwent the ramp-up in frequency, the carrier-phase flow was measured using the line-PIV approach. The carrier-phase streamwise velocity fluctuations were scale decomposed into an extremely large-scale and a smaller scale component. Here, a scale discriminator of $\lambda_x=10\delta$ was used to separate the extremely large-scales from the other scales. These time-dependent extremely large-scale streamwise fluctuations $u_L$ at the near-wall peak, $z^{+}\approx15$ are presented in Fig.~\ref{fig:FluctuationPIV}. The number of mobile particle clusters $N_g$, as the oscillating airfoil underwent a ramp-up in frequency, is also shown in Fig.~\ref{fig:FluctuationPIV}. Note that this has been shifted by 28$tU_{\infty}/\delta$ to account for the downstream convection of flow structures from the oscillating airfoil. Measurements corresponding to three different realizations from identical experiments are presented. In each case, the airfoil oscillation was ramped up from 0 to 15 Hz. For purposes of clarity and visualization, the window of time plotted was restricted to the initial period when the airfoil began to oscillate. Note that the streamwise velocity fluctuations led the particle bed by about $0.28tU_\infty/\delta$, i.e., eddy turnover time scales (refer to Fig.\ref{fig:sch} for line-PIV FOV).

The correlation coefficient defined as $R_{u_L^+N_g}=\mathrm{cov}({u_{L}^{+}N_{g}})/\sigma_{u_{L}^{+}}\sigma_{N_{g}}$ was computed for each of the realizations and is also shown in Fig.~\ref{fig:FluctuationPIV}. Here, $\mathrm{cov}({u_{L}^{+}N_{g}})$ represented the covariance between the extremely large-scale fluctuations in streamwise velocity $u_{L}^{+}$ and the number of mobile particle clusters $N_{g}$, $\sigma_{u_{L}^{+}}$ represented the standard deviation of $u_{L}^{+}$, and $\sigma_{N_{g}}$ represented the standard deviation of $N_{g}$. A strong correlation existed between particle mobilization and these extremely large-scale positive fluctuations. The correlations were all positive ($R_{u_L^+N_g}=0.5$, 0.48 and 0.5 respectively) indicative of the correlation between positive large-scale fluctuations and particle mobilization. There was also some indication that only flow scales which persisted beyond about 50 eddy turnover time scales ($50 tU_\infty/\delta$) are correlated with particle mobilization. This agreed with previous observations from the particle mobilization observed during the ramp-up experiments when particle mobilizations were significant at the early stages of the ramp-up process. As the oscillations progressed in time, the oscillation frequency increased while the number of particles mobilized decreased (see Fig.~\ref{fig:ContinousRampUp}). This may have been attributed to the depletion of the particle bed. However, these coupled line-PIV experiments suggested that this may be the result of the size of the flow scale being introduced. In other words, only very large-scale structures may play a meaningful role in particle mobilization, which is the focus of ongoing work. The current approach also allows for the introduction of energetic flow structures at these scales.

\subsubsection{Oscillation Profile B}

The observation that particle mobilization was correlated with large-scale velocity fluctuations was further investigated by considering isolated (non-periodic) motions of the airfoil and its effect on particle mobilization. During this set of experiments, the airfoil underwent oscillations following Profile B (see Table~\ref{tab:OscProf}). The airfoil was pitched down, followed by a period of rest. After this resting phase, the airfoil was then pitched up and once again held at rest for a certain time period. This process was repeated two more times while keeping the airfoil pitch-up and pitch-down angles constant at $A_f=2^\circ$. However, to isolate the effects of each stroke (pitch-up and pitch-down), the rest period between each successive motion was varied. Note that the speed of the airfoil during each isolated motion was maintained constant at 377~rad~s$^{-1}$, which was the maximum allowed speed of the stepper motor used.

\begin{figure*}[htpb]
\centering
\includegraphics{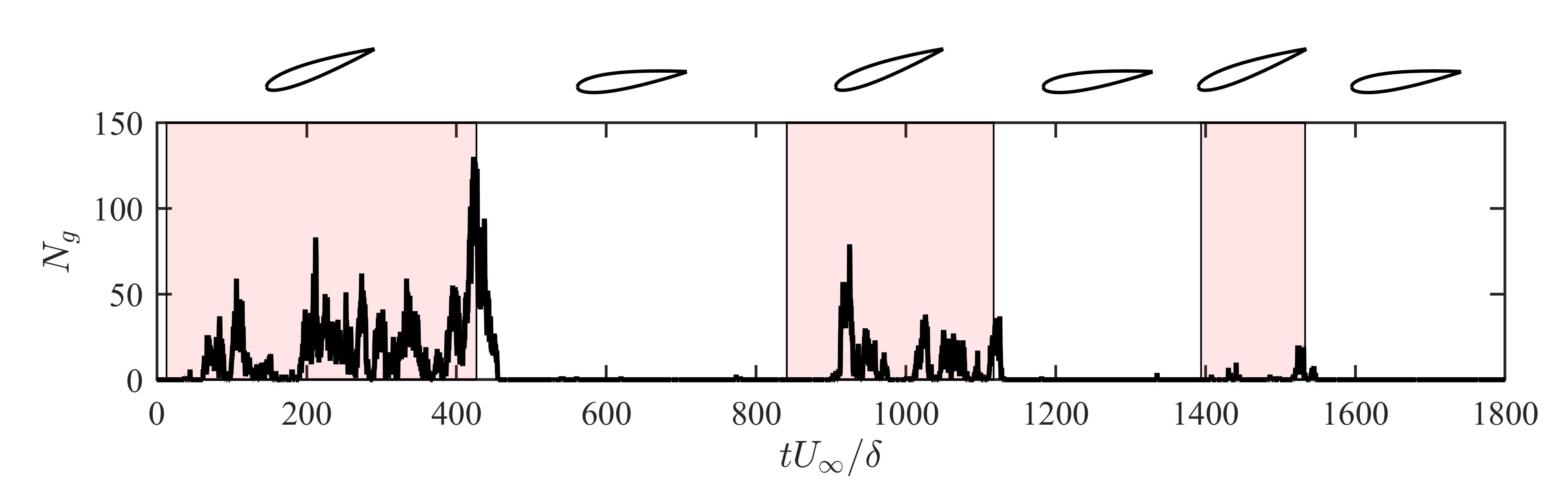}
\caption{Particle motion over oscillation Profile B. The sahded region indicates pitch-down state of the airfoil}
\label{fig:SinglePitchUp}
\end{figure*}

The variation of the mobile particle clusters over time, as the airfoil underwent one such series of motions, is shown in Fig. \ref{fig:SinglePitchUp}. The pink shaded regions represented the time period during which the airfoil remained at rest in the pitch-down position, while the unshaded white regions represented the time spent by the airfoil in the pitch-up position. Observed are periods of high mobilization activity followed by periods of near-zero mobilization. Similar to the observations from Fig. \ref{fig:ContinousRampUp}, these periods of significant particle mobilization were correlated with the individual strokes of the airfoil. Particle mobilization occurred when the airfoil is at rest in the pitch-down position, and this mobilization activity continued until the airfoil returned to its pitch-up position. It was also observed that there was little to no mobilization when the airfoil was in the pitch-up position. The line-PIV measurements depicted in Fig.~\ref{fig:PIVSinglePitchUp} indicated that the airfoil's pitch-down position was associated with high-speed carrier-phase flow in the wall region. Hence, it is concluded that this near-wall increase in momentum, when the airfoil is pitched down, led to the observed increase in particle mobilization. This behavior is consistent with the conclusions drawn from the scale-decomposed line-PIV measurements presented in Fig.~\ref{fig:FluctuationPIV}, where particle mobilization was correlated with large-scale carrier-phase flow structures near the wall. 

\begin{figure}[htpb]
\centering
\includegraphics{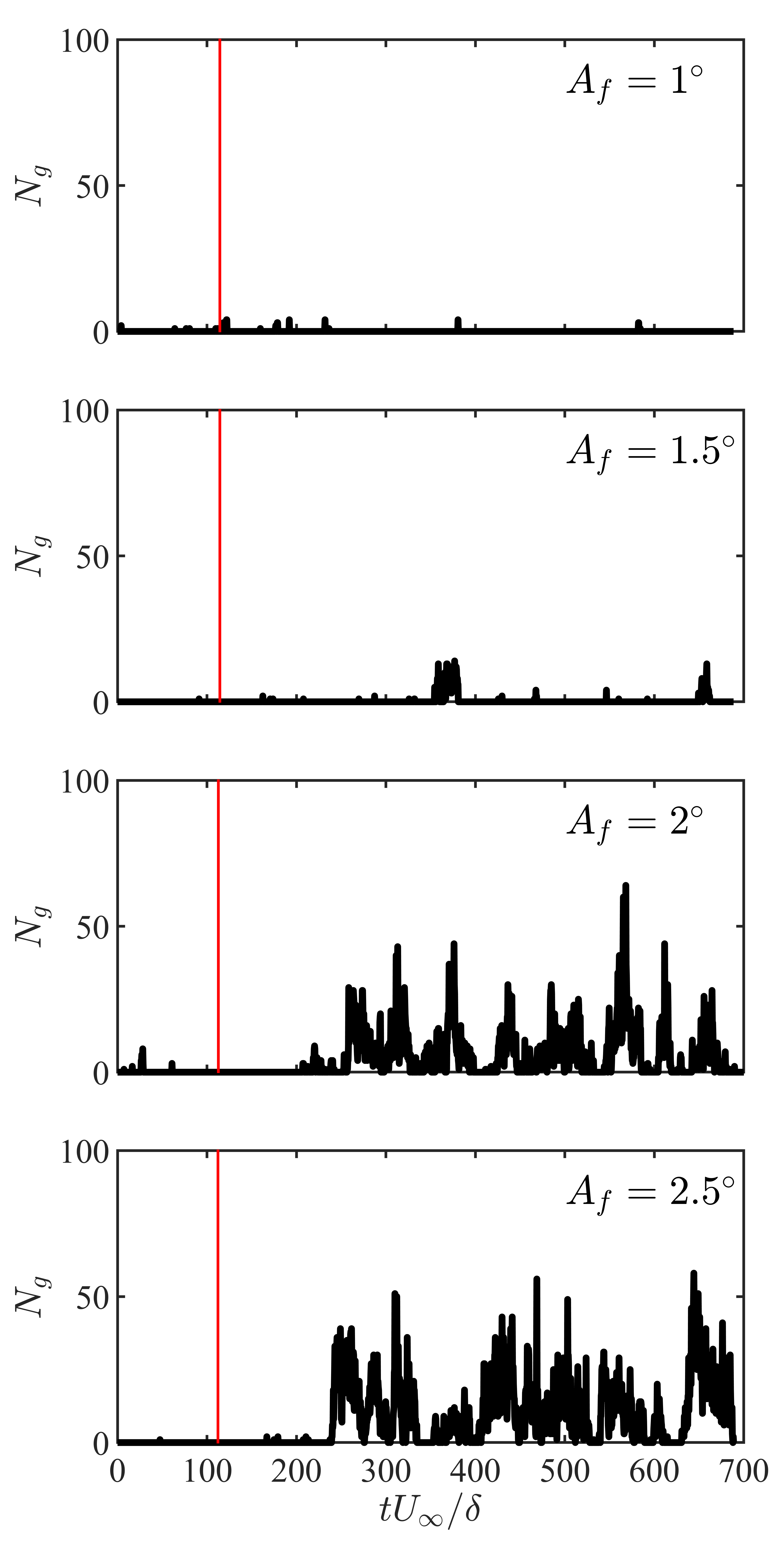}
\caption{Particle mobilization over oscillation Profile C. The instant of pitch-down of the airfoil is indicated by the red line}
\label{fig:Amplitude}
\end{figure}

\subsubsection{Oscillation Profile C}

In order to further demonstrate the capabilities of the current experimental framework, the third oscillation profile, referred to as Profile C in Table~\ref{tab:OscProf} was considered. An isolated pitch-down motion of the airfoil was examined at various amplitudes of oscillation $A_f$. The rationale for considering an isolated pitch-down motion stemmed from inferences drawn from the particle mobilization behavior associated with Profile B. It was noted (from Fig. \ref{fig:SinglePitchUp}) that the pitch-down motion of the airfoil resulted in a period of high streamwise velocity (relative to the local mean) in the near-wall region, which served as the driver of particle mobilization. Hence, Profile C consisted solely of isolated pitch-down motions. The varying amplitudes then represented differing levels of energy introduced into the flow.

The particle mobilization activity as the amplitude $A_f$ of an isolated downstroke of the airfoil is changed is presented in Fig. \ref{fig:Amplitude}. The vertical red line in each subplot indicates the instant at which the airfoil was pitched down. When the amplitude was $A_f=1^\circ$, there  was very little particle mobilization. As the amplitude of pitch-down $A_f$ increased, an increase in particle mobilization occurred. In particular, a large increase in particle mobilization occurred when the amplitude increased from $A_f=1.5^\circ$ to $A_f=2^\circ$. This suggested that a certain threshold in energy must be crossed prior to the initiation of particle mobilization. From another perspective, the energy transferred to the particle bed, because of the energetic large-scale events, must exceed some minimum work required for particle entrainment. These observations are in agreement with previous studies~\citep{Phatz2020a,Valyrakis2013a,Kok2012a,Lee2012a}, which suggest that the initiation of aeolian particle transport is tied to the crossing of a critical energy or work done threshold. More interestingly, the present measurements suggest that the oscillating airfoil mechanism used presents a technique to mobilize particles on a bed ``on demand'', i.e., particle mobilization can be turned on and off. This offers several possibilities in terms of controlling particle mobilization for both controlled processes and carefully controlled studies.

\section{Conclusions}

The controlled periodic oscillation of a NACA-0010 airfoil generated synthetic large-scale structures in a canonical wall-bounded turbulent boundary layer. The oscillation mechanism utilized enabled independent control of both the scale size and the energy of the introduced scales by varying the frequency and amplitude of oscillation, respectively. The flow behind this oscillating airfoil mechanism was carefully characterized while varying both the forcing wavelength and the forcing amplitude. Following this, demonstrative dual-phase measurements were conducted by exposing a particle bed comprising nominally spherical soda-lime particles to this flow developing downstream of the oscillating airfoil. The scale-size and energy was systematically varied using the different oscillation profiles outlined in Table \ref{tab:OscProf}. The primary conclusions are summarized below. 

\begin{enumerate}

    \item Around the near-wall peak, the profiles of the mean velocity and the streamwise turbulence intensity collapsed with that of the canonical flow. This indicated that near-wall production cycle was not interrupted in any significant manner by the forcing. Further away, from the wall into the wake region clear changes in the turbulence intensity and mean velocity profiles were seen. In particular, the oscillations of the airfoil, as expected, increased the free-stream turbulence levels. 
    
    \item The spectral signature around the near-wall peak confirmed that the near-wall cycle was not interrupted in any significant manner. Further, the spectra revealed that the boundary layer responded to the introduced tonal disturbance by increasing the energy associated with those streamwise modes corresponding to the forcing frequency. This was the linear response of the flow. Non-linearities in the form of higher harmonics developed particularly in the case of the larger forcing wavelengths. However, irrespective to the forcing wavelength, a increase in energy at the forcing wavelength was observed in the near-wall region. Further, as the amplitude of oscillation of increased the excess energy at the forcing scale also increased.

    \item Examination of the phase-locked velocity profiles showed the presence of alternating high-speed and low-speed events (relative to the local mean). These alternating events were observed to be correlated with specific motions of the airfoil. In other words, the pitch-down motion of the airfoil resulted in a high-speed event near the wall. Similarly, the pitch-up motion of the airfoil resulted in a low-speed event near the wall. 
    
    \item These observations collectively demonstrated that the oscillations of the airfoil generated a flow field in which a controllable large-scale structure was introduced into a nominally canonical wall-bounded turbulent boundary layer. These scales are controllable in the sense that changing the frequency of oscillation changed the forcing wavelength while changing the amplitude of oscillation changed the energy in these scales.
    
    \item In flow scenarios where substantial natural particle mobilization was absent, the oscillation of the airfoil induced considerable particle mobilization. Particle mobilization was noted to be highly intermittent in nature. This intermittency was positively correlated with extremely large-scale positive fluctuations of the streamwise velocity. In other words, it was observed that particles were preferentially mobilized when a large-scale high-momentum structure passed over the particle bed.  

    \item Isolated pitch-down and pitch-up motion of the airfoil at a constant amplitude were considered. It was observed that high particle mobilization activity immediately followed a pitch-down motion and little to no mobilization activity followed a pitch-up motion. The velocity measurements indicated that the pitch-down motion of the airfoil was associated with increased momentum streamwise flow closer to the wall in agreement with prior observations.

   \item A systematic variation in the amplitude of the pitch-down motion demonstrated a progressively increasing mobilization activity when the amplitude was increased. Given that the amplitude of oscillation was correlated with the energy of the introduced scales, it was concluded that the energy in the introduced scale must surpass a certain threshold to initiate particle mobilization.

\end{enumerate}

\vspace{1cm}
\small
\noindent\textbf{Authors' Contributions} The research was conceptualized by EPG and ZZ. RHB and VT designed the experimental setup under the guidance of EPG and ZZ. Measurements were conducted by VT and RHB. The manuscript was primarily prepared by VT with assistance from RHB and EPG. Finally, the manuscript was critically reviewed and edited by EPG and ZZ.

\begin{acknowledgements}
This research was sponsored by the Army Research Office and was accomplished under Grant Number W911NF-22-1-0253. The views and conclusions contained in this document are those of the authors and should not be interpreted as representing the official policies, either expressed or implied, of the Army Research Office or the U.S. Government. VT would like to thank Kaitlyn Williams for assistance in carrying out measurements.

\end{acknowledgements}

\section*{Declarations}

\noindent\textbf{Conflict of interest} The authors declare no competing interests.

\noindent\textbf{Ethical approval} Not applicable.

\normalsize

\bibliographystyle{spbasic}      
\bibliography{expFluidsBib}   

\end{document}